\documentclass[english,showpacs,amsmath,amssymb,superscriptaddress,prb,twocolumn]{revtex4-1}

\usepackage[T1]{fontenc}
\usepackage[latin9]{inputenc}
\usepackage{color}
\usepackage{amsmath}
\usepackage{graphicx}
\usepackage{amssymb}
\usepackage{esint}

\usepackage{amsmath,amsfonts,amssymb,amstext}
\usepackage{graphicx}
\usepackage[T1]{fontenc}
\usepackage[latin9]{inputenc}
\usepackage{amsmath}
\usepackage{graphicx}
\usepackage{amssymb}
\usepackage{esint}

\makeatletter
\@ifundefined{definecolor}
 {\usepackage{color}}{}

\@ifundefined{definecolor}
 {\@ifundefined{definecolor}
 {\usepackage{color}}{}
}{}
\usepackage{feynmp}
\usepackage{wrapft}% Include figure files
\usepackage{dcolumn}% Align table columns on decimal point
\usepackage{bm}% bold math

\usepackage{babel}

\newcommand{\ket}{\rangle}
\newcommand{\bra}{\langle}
\newcommand{\up}{\uparrow}
\newcommand{\down}{\downarrow}
\newcommand{\beq}{\begin{equation}}
\newcommand{\eeq}{\end{equation}}
\newcommand{\barr}{\begin{array}}
\newcommand{\earr}{\end{array}}

\newcommand{\ed}{\epsilon_d}

\newcommand{\Tr}{\mathrm{Tr}}
\newcommand{\intff}{\int_{-\infty}^\infty}

\newcommand{\intf}{\int_0^\infty}

\usepackage{babel}

\begin{document}
\title{Noise and microresonance of critical-current\\ in Josephson junction induced by Kondo trap states}

\author{M. H. Ansari}
\email{mhansari@uwaterloo.ca}

\affiliation{Department of Combinatorics and Optimization and Institute
for Quantum Computing, University of Waterloo, 200 University Ave.
W, Waterloo, ON, N2L 3G1}
\affiliation{Department of Physics and Astronomy and Institute for
Quantum Computing, University of Waterloo, 200 University Ave. W,
Waterloo, ON, N2L 3G1}

\author{F. K. Wilhelm}
\altaffiliation{present address: Theoretical Physics, Saarland University, 66123 Saarbr\"ucken, Germany}
\affiliation{Department of Physics and Astronomy and Institute for
Quantum Computing, University of Waterloo, 200 University Ave. W,
Waterloo, ON, N2L 3G1} 
%\email{fkwilhelm@gmail.com}
\date{\today}

\begin{abstract}
We analyze the impact of trap states in the oxide layer of a superconducting tunnel junctions, on the fluctuation of the Josephson critical-current, thus on coherence in superconducting qubits.  Two mechanisms are usually considered: the current blockage due to repulsion at the occupied trap states, and the noise from electrons hopping across a trap. We extend previous studies of noninteracting traps to the case where the traps have onsite electron repulsion inside one ballistic channel. The repulsion not only allows the appropriate temperature dependence of $1/f$-noise, but also the control of the coupling between the computational qubit and the spurious two-level systems inside the oxide dielectric.  We use second-order perturbation theory, which allows to obtain analytical formulas for the interacting bound states and spectral weights, limited to small and intermediate repulsions. Remarkably, it still reproduces the main features of the model as identified from the numerical renormalization group. We present analytical formulations for the subgap bound-state energies, the singlet-doublet phase boundary, and the spectral weights. We show that interactions can reverse the supercurrent across the trap. We finally work out the spectrum of junction resonators for qubits in the presence of onsite repulsive electrons 
and analyze its dependence on microscopic parameters that may be controlled by fabrication.
\end{abstract}


\maketitle

%%%%%%%%%%%%%%%%%%%%%%%%%%%%%%%%%%%%%%%%%%%%%%%%%%%%%%%%%%%%%
%%%%%%%%%%%%%%% I N T R O D U C T I O N %%%%%%%%%%%%%%%%%%%%%%%%%%%%%
%%%%%%%%%%%%%%%%%%%%%%%%%%%%%%%%%%%%%%%%%%%%%%%%%%%%%%%%%%%%%

\section{Introduction}

Josephson junction circuits are promising candidates for the physical
implementation of quantum computing. They promise scalability and integrability
from the very start. ~\cite{Makhlin01,You05b,Insight} The grand challenge of realizing quantum computing
in such a mesoscopic solid-state device is decoherence \cite{Nato06II}. In the first
decade of experimentation in this area, decoherence from the electromagnetic
environment has been understood and exquisitely controlled \cite{EPJB03,Martinis03,Makhlin01}. Over the
last years, it has become increasingly clear that the limits in coherence
times are set by properties of the material comprising the Josephson
circuit, in particular its surfaces and interlayer dielectrics. \cite{Harlingen04,Simmonds04,Galperin05,Wang09b,Song09,OConnell08,Astafiev06,Astafiev04b,Kakuyanagi04,Ithier05,Martin05,Shnirman05,Khalil10,Eroms06,Martinis09,Kline09,Oh06,Sendelbach08,Wellstood04,Choi09} 
One prominent source of decoherence is critical-current noise. This noise
has already been discussed in the context of optimizing high-resolution
measurements in superconducting quantum interference devices (SQUIDs). Several models have been put forward, including local moments from
glass physics \cite{Constantin07,Martin05,Shnirman05}, Kondo and other impurities \cite{{Faoro05},{Faoro06},{Faoro07}} and non-interacting Andreev trap states \cite{PRL05,deSousa09}. 

 A typical Josephson junction,  a thin, in most cases non-crystalline, insulator of oxide layer between two pure superconductor reservoirs, may have a fluctuating state that causes subsequent fluctuations in the critical-current.  One can argue these fluctuations could be due to defects in the amorphous material of oxide layer. A single defect, i.e., a two-level system, can generate a Poissonian distribution of current noise, the so-called ``random telegraph noise'', and a bath of these fluctuators with natural energy level distribution may superpose into a non-Gaussian $1/f $ current noise.  The $1/f $ noise has been experimentally observed in the Josephson junction\cite{{Astafiev04,Astafiev06},{Martinis04}}.

There are two experimental results that  illustrate the characteristics of  two-level systems generating this noise.  First, the $1/f $ noise in a junction below the frequency 10Hz displays  a $T^{2}$-dependence on temperature.\cite{Wellstood04}  Second, a more precise measurement performed by Simmonds \emph{et.al.}\cite{Simmonds04}, and subsequently, many others, shows that a current-biased phase qubit during driving transitions between states $|0\rangle$ and $|1\rangle$ is partially blocked at some specific frequencies. This indicates the existence of microwave resonators in the current noise spectrum on top of the $1/f$  spectrum.  These resonances were reported in Ref. \onlinecite{Oh06} to be highly suppressed by $80\%$  if a single-crystal $\textup{Al}_{2}\textup{O}_{3}$ replaces the amorphous tunnel barriers. Nevertheless the nature of this noise and the reason for its enhancement in crystalline dielectrics is left unknown. 
 
Faoro and Ioffe in Ref. \onlinecite{Faoro07} hypothesized that the $T^{2}$-dependence of the $1/f$ -noise sector is satisfied if 1) the two-level systems are shallow Kondo traps, and 2) the majority of these traps have bound states very close to the Fermi surface.  Inside a tunneling junction there are typically a few $\textup{O}_{2}$ impurities left from the oxidization. These  Kondo traps are  impurities with a singly occupied level that carry a free spin. These magnetic impurities can generate  localized bound states
below the superconducting gap. Two potential candidates for generating noise at very low energy ($f<10$ Hz) are (1) slow electrons
and (2) nuclear spins; however  due to the
universal $1/f$  noise persistence only up to 10Hz, which is two orders of magnitude smaller than the typical energy scale associated with nuclear spins energies, \cite{faoro2008, Yoshihara2010, Yoshihara2006} the nuclear spins cannot play a crucial rule in generating this noise. Recent experiments
\cite{Sendelbach2008, Bluhm2009} as well as a density functional
calculation \cite{Choi2009} confirmed the nature of localized electronic
levels to act as electronic spin-oriented traps. 

Constantin and Yu in Ref. \onlinecite{Constantin2007} studied the critical-current noise from the interaction of conduction electrons with a magnetic dipole localized inside the Josephson junction and showed that it is suppressed as the magnetic dipole is located deeper inside the junction.  A deeper study of this phenomenon through quantum field theory unveil more details for the cases of shallowly or deeply located magnetic impurities inside the junction.  If the Kondo traps are coupled to only one of the superconducting reservoirs (i.e., located close to one of the superconductor-insulator interfaces), they may affect the critical-current by (1)  blocking out a part of supercurrent  and (2) generating microwave resonances from the coupling between conduction electron and the electron captured by the trapping center. In a single trap and for the case of non-interacting electrons this has been studied by de Sousa\emph{et al.}\cite{deSousa} where microresonators appear at frequencies twice as the Andreev bound-state energies. However, that study suffers from the absence of Kondo temperature as the electrons do not interact. In the first part of this paper, we extend the noise study to nonzero Kondo temperature.   On the other hand, if the Kondo traps are coupled to both superconducting reservoirs (i.e., located deep in the cener of junction) they affect the critical-current by (1) the quasiparticle poisoning of the Josephson current and (2) the current reversal flow. Faoro, Kitaev and Ioffe in Ref. \onlinecite{faoro2008} studied the quasiparticle poisoning contribution to charge and critical-current noise via a simple toy model. In the second part of this model, we study the current reversal flow effect.

 The inclusion of Coulomb interaction alters the Andreev bound states  to form bound states elsewhere inside the superconducting gap.  As Faoro and Ioffe indicated, \cite{Faoro07} onsite  interaction has a major impact on the $1/f$  noise sector such that without it, the temperature dependence of the current noise cannot be achieved with a reasonable density of states.  Given the possibility of low-lying energy levels near the Fermi sea in the presence of interaction, and assuming that the majority of traps are at low-lying energy levels,  the interaction impacts the charge blockade phenomenon too.

 In this paper, we aim to tackle this problem by studying the critical-current noise generated by repulsive electrons. To achieve this goal, one needs precise solutions of the Anderson impurity model in the Josephson junction. As will be described in more detail, the mean-field approximation to these problems often leads to unphysical result, thus one needs to resort to non-perturbative methods such as the numerical renormalization group (NRG) \cite{hechtPaper}, the functional renormalization group (fRG) \cite{Salmhofer01}, and density matrix renormalization group (DMRG) \cite{Schollwoeck05,Saberi08} . These methods have achieved the precise description of the competition between superconducting order and the effect of the impurity. Nevertheless, NRG and DMRG need heavy  numerical analysis and fine-tuning of numerical parameters, and fRG suffers from many truncations necessary to numerically generate the appropriate self-energies. Unfortunately, there is no secure analytical solution that is as reliable as NRG and that accurately describes experiments. We  study this problem analytically within second-order  perturbation theory. We derive the analytical weights associated with the subgap resonances in the interacting theory and interestingly reproduce some of the features of interacting theory known previously by numerical analysis of NRG and fRG. After a discussion of the range of validity of our approach, we highlight the noise spectrum due to blockade of current by occupied traps.

%%%%%%%%%%%%%%%%%%%%%%%%%%%%%%%%%%%%%%%%%%%%%%%%%%%%%%%%%%%%%
%%%%%%%%%%%%%%% INTRO: ..... PI JUNCTION %%%%%%%%%%%%%%%%%%%%%%
%%%%%%%%%%%%%%%%%%%%%%%%%%%%%%%%%%%%%%%%%%%%%%%%%%%%%%%%%%%%%

\subsection{$\pi$-junctions and interaction effects}

A Josephson junction is characterized by a supercurrent $I(\phi)$
that is an antisymmetric, $2\pi$-periodic function of the phase
difference between the superconductors \cite{Golubov04}. In the regime of weak and time-reversal-invariant coupling between  superconducting reservoirs across the junction, e.g., in the nonmagnetic
tunnel junctions, we have the 
dc Josephson current $I=I_{c}\sin\phi$ with  the phase
difference of order parameter of the two superconducting reservoirs
connecting via a junction and $I_{c}$ the critical-current \cite{Josephson62}.

The electron transport through a junction between two superconducting
reservoirs is heavily affected in the presence of magnetic impurities.
\cite{{Abrikosov61},{matsuura}} This coupling breaks the time-reversal
symmetry and leads to the appearance of states below
the superconducting energy gap inside the junction.
When the ground state is occupied by one electron, another electron
is repelled from it.  This repulsion affects the current flow such that the conduction electrons may either screen the trapped electron or flip its spin.  This is determined through the  competition between electronic interaction (characterized by  Kondo temperature $T_{K}$) and superconductivity (characterized by the superconducting gap $\Delta$). The possibility of spin flip scattering, no matter if it is a dominant process or not, results in the modification of the bound-state energies such that it becomes possible that at a specific interaction strength the bound states cross the Fermi surface.

In a metallic system the Kondo traps occupied by  one electron spin at high temperature causes conduction electrons to scatter off inelastically in a spin flip process, while at low temperature, they are screened forming a bound singlet state leaving only elastic scattering.  The Kondo temperature $T_K$ is the temperature crossover between the two different scatterings.   However, in a superconducting system more complication appears due to the competition between the Kondo temperature and the superconducting gap.   When the repulsion between onsite electron and the conduction electrons is weak (i.e., large Kondo temperature, $T_{K}\gg \Delta$) in the superconducting phase the magnetic moment  is screened by the conduction electrons such that transport through the junction occurs without spin flip and the ground state of the system (superconductor + impurity) is singlet. In this case the electron at the magnetic impurity along with the conduction electrons form a bound-state below the superconducting gap at the impurity site.  However, strong interactions (i.e., low Kondo temperature, $T_{K}\ll \Delta$) causes the conduction electrons to spin flip at the impurity site, where the ground state of the system is a doublet. This affects the dc current to flow opposite to the phase drop, i.e., $\left.dI(\phi)/d\phi\right|_{\phi=0}<0$. This sometimes is referred to as a $\pi$-junction, as in this case the ground state at zero phase bias is characterized by $\phi=\pi$.  In general, the resulting current-phase relation is not a simple sinusoid. There is a ``certain'' Kondo temperature that serves as the border between spin flip phenomenon and the screening. Recently, using NRG method, this divider Kondo temperature was reported to be $T_{K}^{*}\approx 0.3 \Delta$.\cite{bauer}

\begin{figure}[h]
 \includegraphics[width=7cm]{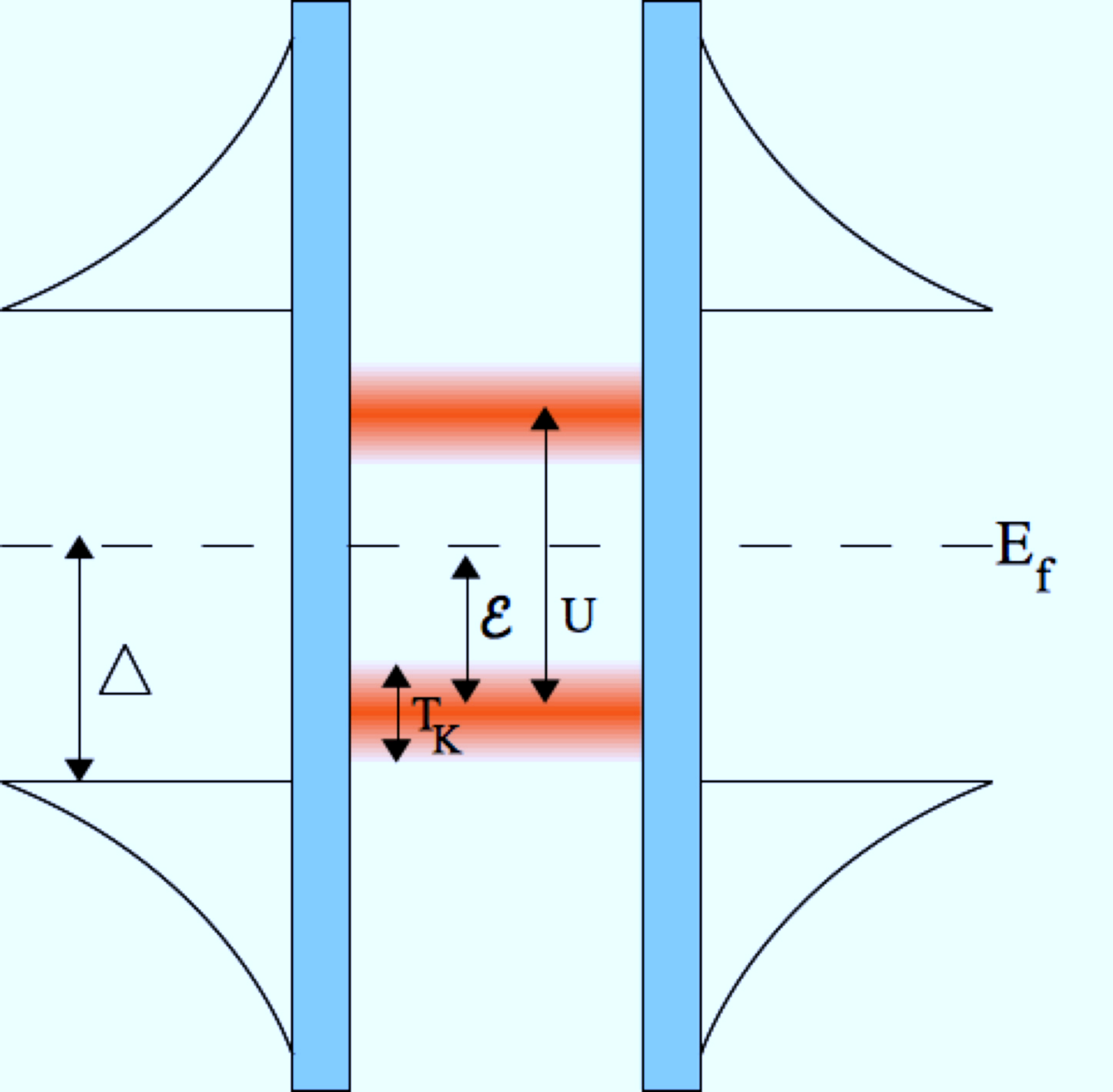}{(a)}\\ \includegraphics[width=7cm]{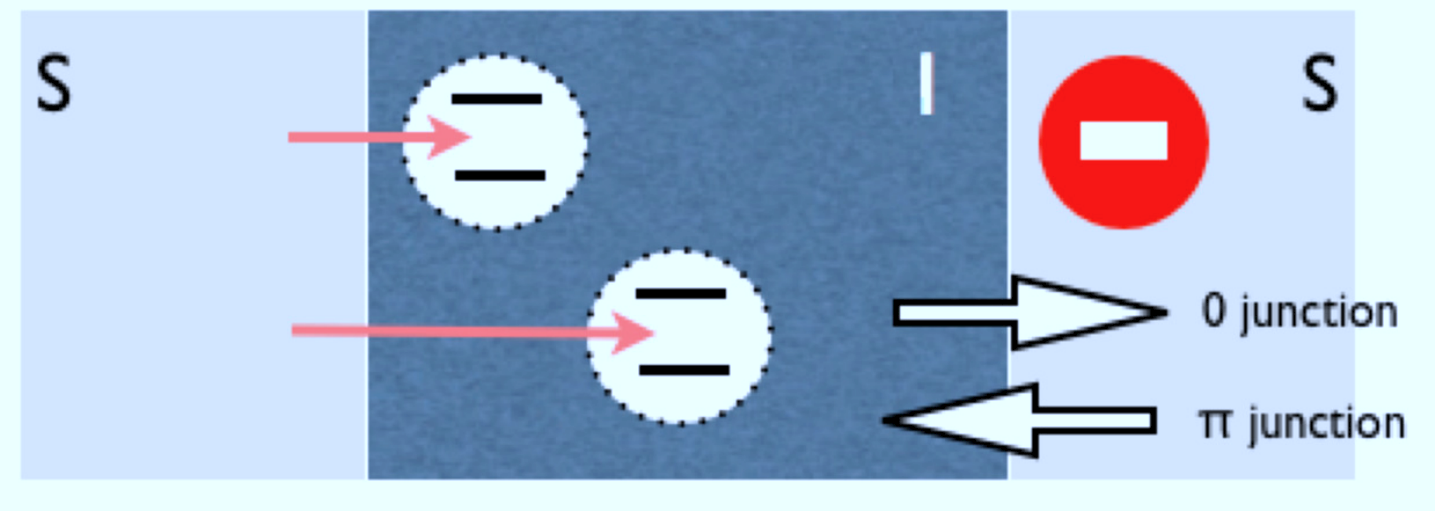}{(b)}  
 \caption{(Color online)  (a)  Energy view on a Josephson junction sandwiched between two superconducting
reservoirs. The two thick (red color) bands are the subgap bound states caused by the presence of magnetic impurity in the junction. (b) Cross-section of the tunneling layer. Traps near the oxide surface (left) predominantly contribute to blocking noise ---an occupied trap is charged and blocks off parts of the current. Traps in the center carry a contribution to the supercurrent that is influenced by interactions and can potentially lead to a $\pi$-junction. }
\label{fig. schematic} 
\end{figure}

There have been various proposals and observations of $\pi$-junctions including those at high-temperature superconductors, non-equilibrium mesoscopic junctions, and magnetic junctions. Here, we are discussing an interaction-induced mechanism. When two superconducting reservoirs are connected via a Josephson junction whose width is sufficiently smaller than the coherence length, 
spin flip tunneling processes can cause supercurrent reversal \cite{shibaSoda,glazman89,spivak91}. More recently van Dam and Nazarov observed this phenomena in a superconducting quantum dot \cite{vanDam:2006fj}. It was shown that at high Kondo temperature (i.e $\Delta \ll T_{K}$) the Josephson current through a Kondo impurity is non-sinusoidal. Ishizaka calculated the Josephson current numerically within the non-crossing approximation; 
it was reported that as a function of the Coulomb repulsion the supercurrent is suppressed
until at some critical value a sudden sign change occurs \cite{Ishizaka95,Shimutzu98} .
Within the numerical renormalization technique, it was reported  that a $\pi$-junction is the preferred state in junction when the impurity is singly occupied and the onsite Coulomb interaction is large \cite{RozhkovArovas,yoshiokaOhashi,matsumoto2000}. A generalization of the non-crossing approximation found a modification in this $0$ to $\pi$ junction transition at the presence of bound states \cite{ambegaokarclark}.  Recently some other
approaches including  NRG method\cite{{choi},{karrasch}}, and fRG \cite{karrasch} were taken to study this effect on a quantum dot and they predicted
a 0-$\pi$ transition at large Coulomb repulsion.

  Figure (\ref{fig. schematic}(a)
illustrates bound-state energies ($-\varepsilon$ and $U-\varepsilon $) inside the Josephson junction caused by magnetic impurity.  In the particle hole symmetrically
case the two states are located symmetric above and below the Fermi surface, i.e., $\varepsilon:= U/2$)  The renormalized  width of these levels is determined by the Kondo temperature.  As indicated by Yoshida and Ohashi \cite{yoshiokaOhashi}
and confirmed by Bauer
et. al.\cite{bauer} the Kondo temperature in the symmetric case is defined by $T_{K}=0.182U\sqrt{8\Gamma/\pi U}\exp(-\pi U/8\Gamma)$, with $U$ being the Coulomb repulsion and $\Gamma$ being the transition rate between superconductor and impurity.

\begin{figure*}[!ht]
\includegraphics[width=9cm]{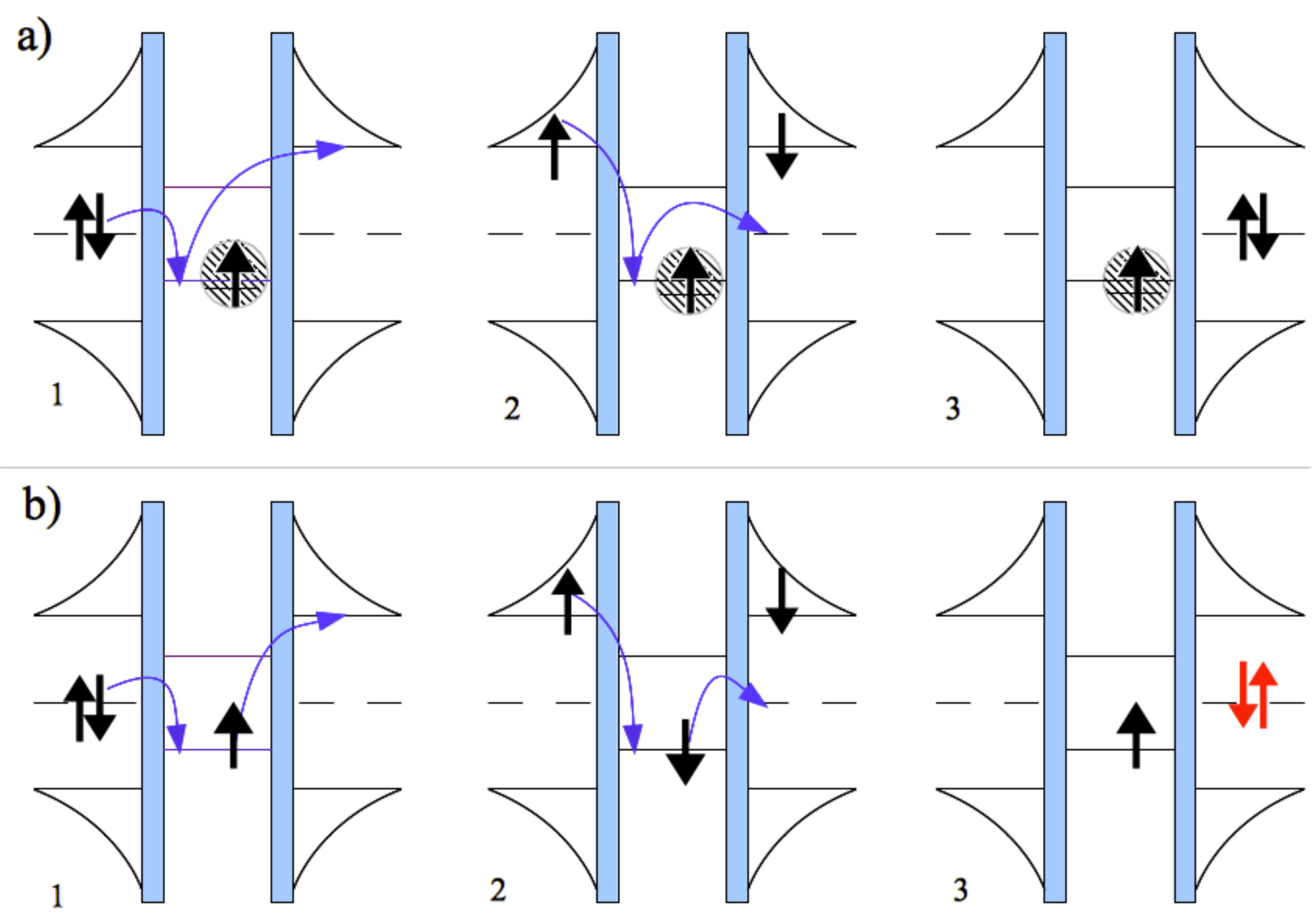}  
 \caption{(Color online)  Energy diagrams illustrating Cooper pair transport through a magnetic impurity. (a) Weak Coulomb-interaction limit where transport occurs through a screened spinless level; denoted by screened arrow at the bound-state $E_{b}$. The Cooper pair parity is conserved.  (b)  Strong- Coulomb-interaction limit where transport flips the impurity spin. The parity of the Cooper pair is reversed, which results in a reversal supercurrent flow.}
\label{fig. schematic2} 
\end{figure*}

 Figure (\ref{fig. schematic2}) shows transport inside a junction in the presence of a magnetic impurity. Let us first consider the first row of panels (a1)--(a3). In weak-interaction limit, the impurity electron is screened by  conduction electrons, therefore the other electrons are scattered only from the potential of this singlet and no spin exchange occurs at the impurity site.  The panel a1 shows the broken Cooper pair in the left reservoir due to proximity effect inside the insulator junction and either one of the two electrons can tunnel to the other reservoir [see panel (a3)] via the virtual state of panel (a2) without spin exchange. The two electrons after consecutive tunneling reproduce the original Cooper pair now in the right side with parity conservation. In the strong repulsion regime (i.e., $U>U_{c}$), the impurity electron is unscreened and the conduction electrons caused by breaking Cooper pairs inside the junction exchange their spin with the impurity electron. As it is shown in the bottom row of Fig. (\ref{fig. schematic2}) panels (b1)--(b3), due to the Pauli principle between the two outcome electrons the one with opposite spin first occupies the impurity level and the impurity electron may tunnel out. Repeating the same process causes a spin flip on both electrons reproducing a Cooper pair in the other reservoir.  In other words, the initial Cooper pair state $(|\up \down \rangle - |\down \up \rangle)/\sqrt{2}$ after passing through this junction becomes $e^{{i \pi}}(|\up \down \rangle - |\down \up \rangle)/\sqrt{2}$.  This simply results into a $\pi$ shift in the Josephson relation and a negative sign for the supercurrent.

%%%%%%%%%%%%%%%%%%%%%%%%%%%%%%%%%%%%%%%%%%%%%%%%%%%%%%%%%%%%%
%%%%%%%%%%%%%%% INTRO: ....  OCCUPATION BLOCKADE %%%%%%%%%%%%%%%%
%%%%%%%%%%%%%%%%%%%%%%%%%%%%%%%%%%%%%%%%%%%%%%%%%%%%%%%%%%%%%

\subsection{Occupation-blockade noise spectrum}

A more conventional junction noise mechanism is based on the occupation
number of the trap state. The idea is that whenever the trap is occupied
by one or two electrons, Coulomb repulsion originating from the localized
trap will block off part of the junction area, hence influencing other
channels. This requires the trap to be coupled to only one of the
electrodes, whereas for the $\pi$-junction scenario we require that
the coupling of the trap to both electrodes is strong enough. Occuption
noise has been studied early in the 1980s \cite{{Wakai87},{Weissman}} and more recently  \cite{{Faoro05},{Faoro06},{Faoro07},{deSousa},{Shnirman}, {Shnirman1}}. These mechanisms are illustrated in Fig. (\ref{fig. schematic2}).

 The charge-blockade decoherence determines how sensitive the microwave resonances, present on top of $1/f$  noise spectrum\cite{simmonds}, are to the Coulomb repulsion. As a result, both the energy of these resonances and their intensities are heavily affected in strong Coulomb repulsion.

 Our goal is to use an analytical approach to the problem. It is well-known that the Kondo effect
in normal metals cannot be described by a mean-field study because it is uncontrolled and underestimates correlations \cite{{anderson},{vondelft}}; however
when this method was utilized by Yoshioka and Ohashi \cite{yoshiokaOhashi}
on superconductors interestingly it was shown at a critical Coulomb interaction the
current starts to flow backward, which means some $0-\pi$ transition is captured
in mean field limits. Nevertheless, later on it was understood this is the artifact of an unphysical effect because in the mean field approximation there is no well defined Kondo temperature that allows to control the phase transition from a non-magnetic
phase into a magnetic one and this current reversal flow appears only
because of the unphysical local moments \cite{karrasch}.

 We extend this study into the second-order  perturbation in Coulomb repulsion and will show that the results are remarkably consistent with the recent non-perturbative results taken from NRG\cite{{bauer},{karrasch}}. Using the equilibrium Green's function method in Sec. II, we  derive an analytic formula for the the dependence of the self-energies in the Coulomb repulsion in Sec. III.  We see that as the Coulomb repulsion increases the subgap states come closer to the Fermi surface and at a critical repulsion $U_{c}$ they overlap it. At stronger repulsion the  states move away from the Fermi surface toward the gap. Bauer \emph{et.al.}\ cite{bauer} employed the NRG method, which is a reliable approach to analyze low temperature physics of spectral functions. The border between singlet and doublet ground states (i.e., $T_{K}/\Delta\approx0.3$) was estimated by the crossover of bound states from the Fermi surface. We reproduce this result and present an analytical formula for this border.   Within the transparent transmission regime, the supercurrent is worked out in Sec. IV and it is shown it changes its flow direction in strong interactions. In the final section, the current noise due to the charge blockade in the junction is worked out and the microresonators are detected for a different Kondo temprature.

%%%%%%%%%%%%%%%%%%%%%%%%%%%%%%%%%%%%%%%%%%%%%%%%%%%%%%%%%%%%%
%%%%%%%%%%%%%%% G E N E R A L    F R A I M W O R K %%%%%%%%%%%%%
%%%%%%%%%%%%%%%%%%%%%%%%%%%%%%%%%%%%%%%%%%%%%%%%%%%%%%%%%%%%%

\section{Single impurity inside the Josephson junction: The general framework}

%%%%%%%%%%%%%%%%%%%%%%%%%%%%%%%%%%%%%%%%%%%%%%%%%%%%%%%%%%%%%
%%%%%%%%%%%%%%% GEN FRAM: .... HAMILTONIAN %%%%%%%%%%%%%%%%%%%%%%%%%%%%%
%%%%%%%%%%%%%%%%%%%%%%%%%%%%%%%%%%%%%%%%%%%%%%%%%%%%%%%%%%%%%

\subsection{Hamiltonian}

Consider two s-wave superconducting reservoirs $j=L,R$, different
only in their phases, described by the mean-field Hamiltonian $\mathcal{H}^{j}:=\sum_{k,s}\epsilon_{k}n_{k,s}^{j}+\sum_{k}\Delta\exp(i\phi_{j})\ c_{k,\uparrow}^{j}c_{-k,\downarrow}^{j}+\mathrm{h.c.}$,
where $n_{k,s}^{j}:=c_{k,s}^{j\ \dag}c_{k,s}^{j}$ is the number of
particles with momentum $k$ and spin $s$ at the reservoir $j$.
One magnetic impurity is inside the junction with the Hamiltonian
$\mathcal{H}^{\mathrm{imp}}:=\sum_{s}\epsilon_{d}n_{s}-Un_{\up}n_{\down}$,
where $n_{s}:=d_{s}^{\dag}d_{s}$ denotes the number of particles
with spin $s$, and the Coulomb repulsion $U$ between electrons.
The coupling between the impurity and the reservoirs is defined by
$\mathcal{H}_{T}^{j}:=t\sum_{k,s}(c_{k,s}^{j\ \dag}d_{s}+\mathrm{h.c.})$
where $t$ is the hopping matrix element. The total Hamiltonian becomes

\begin{equation}
\mathcal{H}:=\mathcal{H}^{{\rm imp}}+\sum_{j=1}^2\left(\mathcal{H}^{j}+\mathcal{H}_{T}^{j}\right)\label{eq:eq H}\end{equation}

We assume the chemical potential is the constant Fermi energy (set as a reference energy), i.e.,
there is no bias voltage. We also consider the two reservoirs
to be identical on all of their parameters except of their phases, 
their phase difference is $\phi_{1}=-\phi_{2}=\phi/2$. We further assume
particle-hole symmetry $\epsilon_{d}=-U/2$. We also define the total
coupling strength to the $j$-th reservoir $\Gamma_{j}:=\pi \rho|t_{j}|^{2}$, where $\rho$ is the density of states.

%%%%%%%%%%%%%%%%%%%%%%%%%%%%%%%%%%%%%%%%%%%%%%%%%%%%%%%%%%%%%
%%%%%%%%%%%%%%% GEN FRAM: ..... GREEN'S FUNCTION  %%%%%%%%%%%%%%%%%
%%%%%%%%%%%%%%%%%%%%%%%%%%%%%%%%%%%%%%%%%%%%%%%%%%%%%%%%%%%%%

\subsection{Green's function}

We are using Green's functions in Nambu space \cite{nambu}, here
and henceforth marked by a hat " $\hat{}$ ". The Green's function corresponding to
the Hamiltonian (\ref{eq:eq H}) in the non-interacting limit at the
impurity site is known \cite{rodero95} to be $\hat{G}_{o}^{d}(\omega)=(\tilde{\omega}-\ed\hat{\sigma}_{z}+\hat{\sigma}_{x}\tilde{\Gamma})^{-1}$,
where $\tilde{\omega}:=\omega(1+\sum_{i}\Gamma_{i}/E(\omega,\Delta_{i}))$
and $\tilde{\Gamma}:=\sum_{i}\Gamma_{i}\Delta e^{i\phi_{i}}/E(\omega,\Delta_{i})$.
The function $E(\omega,\Delta)=-i\ \mathrm{sgn}(\omega)\sqrt{\omega^{2}-\Delta^{2}}$ above the gap is the quasiparticle energy, and 
for the (imaginary) frequencies below the gap as $E(\omega,\Delta)=\sqrt{\Delta^{2}-\omega^{2}}$.
With the choice of identical superconducting reservoirs one can see $\tilde{\omega}=\omega(1+2\Gamma/E(\omega,\Delta))$
and $\tilde{\Gamma}=2\Gamma\Delta\cos(\phi/2)/E(\omega,\Delta_{i})$.

The Dyson-Gorkov equation formally gives the interacting Green's function
in terms of the diagonal and off-diagonal self-energies $\Sigma$
and $\Sigma_{\Delta}$ 

\begin{equation}\label{eq. G}
\hat{G}_{d}=-\frac{1}{F(\omega,\phi)}\left(\begin{array}{cc}
\tilde{\omega}-\epsilon_{d}-\Sigma(\omega) & \tilde{\Gamma}-\Sigma_{\Delta}(\phi)\\
\tilde{\Gamma}^{\ast}-\Sigma_{\Delta}^{\ast}(\phi) & \tilde{\omega}+\epsilon_{d}+\Sigma^{\ast}(-\omega)\end{array}\right)\end{equation}

where $F(\omega)$ is the determinant of the matrix.

%%%%%%%%%%%%%%%%%%%%%%%%%%%%%%%%%%%%%%%%%%%%%%%%%%%%%%%%%%%%%
%%%%%%%%%%%%%%% GEN FRAM: .... GOING BEYOND MEAN FIELD  %%%%%%
%%%%%%%%%%%%%%%%%%%%%%%%%%%%%%%%%%%%%%%%%%%%%%%%%%%%%%%%%%%%%

\subsection{Going beyond mean field}

As one analytical approach, we are choosing second-order  perturbation theory in $U/2\pi\Gamma$.
It will provide analytical expressions for the subgap bound-state energies and their spectral weights, which are otherwise precisely known only 
numerically through NRG results. The lowest order mean field theory, on the other hand, has the known restrictions already discussed. We will see that from finding the second-order  self-energy matrix in Nambu space (and in the particle-hole symmetry), we can extract  analytical formulas for the bound states and their spectral weights in the non-magnetic phase. We noticed these results are in agreement with NRG results and thus reliable for our purpose to study noise in the domain of validity of this perturbative approach.  

The significant improvement from including second-order  perturbative theory at least in the non-magnetic phase relative to mean field theory can be tracked down to a set of reasons.
The self-energy in the particle-hole-symmetric case has (diagonal) particle-particle as well as (off-diagonal) particle-hole matrix elements in Nambu space. In the mean field approximation the Coulomb repulsion modifies the diagonal self-energy by shifting the bare electron and hole energies and the off-diagonal self-energy by inducing an order parameter into particle-hole pairs. This self-energy leads to poles of the Green's function, i.e., induces bound states below the superconducting gap.  The subgap energies were found analytically \cite{matsumoto2000} as well as self-consistently\cite{{shiba73},{yoshiokaOhashi}}.  The second-order  perturbation adds a term that depends on $U^{2}$, $\Gamma$, and $\omega$ into the diagonal self-energies and a term that depends on $U^{2}$, $\Gamma$, and $\phi$ into off-diagonal self-energy. Some of the improvements these terms cause are listed below. 

\begin{enumerate}
\item  The diagonal mean field self-energy is frequency-independent.\cite{matsumoto2000} This leads to a vanishing Kondo temperature $T_{K}$, i.e., it is insensitive to interaction---see next section. Lacking an interaction-controlled  Kondo temperature the singlet-doublet transition, which is supposed to be governed by $T_{K}/\Delta$, becomes uncontrollable. The second-order  self-energy improves this by providing a frequency-dependent self-energy and therefore an interaction-sensitive  Kondo temperature ---see section \ref{sec self-energy}.

\item Accurate NRG results show that, as the repulsion between electrons increases, the bound states come closer to the Fermi surface, and after overlapping it at a critical interaction $U_{c}$ they slowly move away from the Fermi surface. This behavior is captured in the self-consistent mean field solution\cite{yoshiokaOhashi}; however, the criticality appears to be independent of the superconducting gap $\Delta$. This contradicts the precise NRG\cite{bauer} result as $U_{c}$ must decrease for  superconductors with larger gaps. The second-order  perturbation improves this into an NRG-like $\Delta$-dependence of the criticality. This dependence can be seen in  Eq. (\ref{eq. Uc}) below.

\item  The phase boundary between the singlet and doublet ground states in the strong coupling regime was worked out by NRG methods and estimated  by $T_{K}/\Delta \approx 0.3$.\cite{{yoshiokaOhashi},{satori92},{sakai}} This boundary can also be approximated by the criticality line $U_{c}$.\cite{bauer}   However, in the mean field, the Kondo temperature is not tuneable. In the second-order  the border $T_{K}$ is recovered as a an inversely scaling interaction for which this border can be well-approximated by the criticality line ---see Fig. (\ref{fig. phase}).

\item The self-consistent mean field solution describes supercurrent reversal in the strong Coulomb repulsion regime, a $0-\pi$ transition. This transition turns out not to be sensitive to the relevant physical  parameters, in contrast to more accurate solutions suggesting that it should be controllable.\cite{{yoshiokaOhashi},{matsumoto2000},{karrasch}} This sign change is recovered in a controllable way in the second-order  with a modified current-phase relation due to the $\phi$ dependence of the bound-state energies appearing in that order. 
\end{enumerate}

Let us emphasize our solution is analytical and restricted to the non magnetic phase. Also we are not aware of any self-consistent analytical solution in the entire phase diagram.

%%%%%%%%%%%%%%%%%%%%%%%%%%%%%%%%%%%%%%%%%%%%%%%%%%%%%%%%%%%%%
%%%%%%%%%%%%%%% 2 N D   O R D E R  P E R.  %%%%%%%%%%%%%%%%%%%%%%%%%%%%%
%%%%%%%%%%%%%%%%%%%%%%%%%%%%%%%%%%%%%%%%%%%%%%%%%%%%%%%%%%%%%

\section{second-order  perturbation theory}

%%%%%%%%%%%%%%%%%%%%%%%%%%%%%%%%%%%%%%%%%%%%%%%%%%%%%%%%%%%%%
%%%%%%%%%%%%%%% 2ND ORD:  ..... SELF ENERGY  %%%%%%%%%%%%%%%%%%%%
%%%%%%%%%%%%%%%%%%%%%%%%%%%%%%%%%%%%%%%%%%%%%%%%%%%%%%%%%%%%%

\label{sec self-energy}
\subsection{Self-energy }

The second-order  perturbation of the Coulomb interaction in the Nambu space can be computed by Feynman diagrams using normal and anomalous propagators. The diagonal self energy in the first order is known to be\cite{roderoPi} $\Sigma^{(1)}=U/2$, which shifts the bare energy of impurity into $\tilde{\epsilon_{d}}=\epsilon +U/2$. Usually the particle hole symmetry is imposed onto the problem to set $\tilde{\epsilon_{d}}=0$. The off-diagonal self energy in the first order is $\Sigma^{(1)}_{\textup{off}}=-\Delta_{d}$, where $\Delta_{d}$ is the induced superconducting order parameter at the impurity site. By definition, $\Delta_{d}:=U\langle d_{\down}d_{\up}\rangle=\frac{U}{8\pi}\intf d\omega\ \Tr(\sigma_{x}\hat{G}^{d}(\omega))$, which contains the off-diagonal element of the impurity Green's function.

As discussed in the introduction the low energy	physics is sensitive to the Kondo temperature $T_{K}$ which essentially measures the width of the subgap many-body resonance. The Kondo temperature in perturbation theory is defined as $T_{K}\approx \frac{\Gamma}{1- \partial \Sigma / \partial \omega}$ at $\omega=0$.\cite{roderoPi} In the symmetric Anderson model this temperature vanishes in the mean field limit. Therefore, any phase transition in the junction that is governed by the ratio of $T_{K}/\Delta$, if even occurs, is unphysical.

In the second-order , the self-energy is defined as
 $\Sigma(U)=\frac{U}{2}+\alpha(U)$ and $\Sigma_{\Delta}:=\beta(U,\phi)\ \Delta$, where
$\alpha(U)=a(U) \omega $, and $\beta(U,\phi)=\frac{\Delta_{d}}{\Delta}+\beta^{(2)}(U,\phi)$
for $\beta^{(2)}(U,\phi)=b(U) \cos (\phi/2)$.  These formulas are extracted from Feynman diagrams and details are given in Appendix A.  Considering this order of perturbation theory the Kondo temperature not only does not vanish, but also it scales inversely with $U$, which although fails to provide the exact exponential decay, yet supports decaying with increasing Coulomb repulsion.

Note that in the particle-hole symmetric case the Friedel sum rule remains  satisfied.\cite{hewsonBook} .
However, beyond this symmetry a modification is needed on the self-energy in order to satisfy the Friedel sum rule.
A typical modification was proposed by Martin-Rodero and Levy-Yeyati and coauthors \cite{rodero95,roderoPi,roderoSigma,hewsonBook}
 that maintain the sum rule. A brief overview of this modification can be found in the Appendix B. Nevertheless, because in this paper we consider the particle-hole symmetry for granted, the Friedel sum rule is satisfied and no modification is required. 

The second-order  Feynman diagrams  read
\begin{eqnarray}
\Sigma^{(2)}(\omega) &=& - \left(3-\frac{\pi^{2}}{4}\right) \left(\frac{U}{2\pi\Gamma}\right)^{2} \omega,\\
\Sigma^{(2)}_{\textup{off}}(\phi) &=&  \left(  \frac{ \pi^{2}}{6} - 1\right) \left(\frac{U}{2\pi \Gamma}\right)^{2} \Delta  \cos \frac{\phi}{2}.  \\ \nonumber \label{eq. self energies}
\end{eqnarray}

The diagonal self-energy was first reported in 1975 in normal metals by Yoshida and Yamada\cite{{yamada},{yamada1},{yamada2}} in the Anderson model in the normal conducting state.  As discussed earlier in Ref. \onlinecite{roderoPi} and mentioned in Eq. (\ref{eq. App G prop}) the normal propagator in the small-gap limit can be assumed not to be modified by superconductivity.  The superconductivity effect enters the off-diagonal propagator.   Having written the effective Hamiltonian as the summation of non-interacting part and the self-energy interacting part,  the perturbative energy ratio of diagonal Hamiltonian is almost $\sim E_{b}(U)/8\Delta$ in the unit of $(U/ \pi \Gamma)^{2}$. This indicates that the perturbatively relevant limit is  $U/ \pi \Gamma <  2.8\sqrt{\Delta/ E_{b}} $, thus the closer a bound-state is to the Fermi surface the more accurate is the perturbation theory on diagonal interactions. However, for off-diagonal interactions the perturbation limit turns out to be $U/ \pi \Gamma <  2.5$.  It is important to emphasize that the responsible physics can only be inferred within the limit of $\min \{ 2.5,   2.8 \sqrt{\Delta/ E_{b}}\}$. Note that in this section, because we want to compare our results with NRG results, we go a bit beyond the perturbation limit and interestingly reproduce all NRG physics.  Nevertheless we will remain in the perturbative limits for  the rest of this paper where supercurrent reversal and its decoherence are discussed.    

Note that current in a not very long weak link vanishes at the phase $\phi=\pi$ which is also confirmed in experiments.\cite{phase-currentReview} This would not be achieved by the self-energy found in Ref. [\onlinecite{roderoPi}], but by the one found in this paper.

%%%%%%%%%%%%%%%%%%%%%%%%%%%%%%%%%%%%%%%%%%%%%%%%%%%%%%%%%%%%%
%%%%%%%%%%%%%%% 2ND ORD:  .... BOUND STATES  %%%%%%%%%%%%%%%%%%%%%%
%%%%%%%%%%%%%%%%%%%%%%%%%%%%%%%%%%%%%%%%%%%%%%%%%%%%%%%%%%%%%

\subsection{Bound-states}

The magnetic moment at the impurity site is defined by the difference in occupation between the spin polarizations, $m:=(-U/2)(\langle n_{\up}\rangle-\langle n_{\down}\rangle)=-\frac{U}{8\pi}\intf d\omega\ \Tr(\sigma_{z}\hat{G}^{d}(\omega))$. This contains the difference between two diagonal elements of impurity Green's function. This definition describes the spontaneous symmetry breaking responsible for distinguishing magnetic from non-magnetic states. Because of the possible difference between spin polarizations we define $\eta:=(\epsilon_{d}+U/2)-m$ as the (bare) energy of magnetic impurity.  However, we restrict this study to the non-magnetic phase sector, thus $m=0$ everywhere. Due to the particle-hole symmetry the first order diagonal self-energy becomes $U/2$. Therefore in Eq. (\ref{eq. G}) the bare energy $\epsilon_{d}$ is summed up with the first order self-energy to be renamed $\eta$. This shifted energy is now playing role as a new bare energy to be renormalized by the second-order  self energy. 

The subgap states appear where the denominator of the Green's function
in Eq. (\ref{eq. G}) vanishes, i.e., $F(\pm E_{b})=0$. The bound-state energies of these states are the solutions to 

\begin{equation}
\left(\Delta^{2}-E^{2}\right)\left(1-\alpha+\frac{2\Gamma}{E}\right)^{2}-\eta^{2}=\left(\frac{2\Gamma\Delta\cos\frac{\phi}{2}}{E}-\beta \Delta\right)^{2}
\label{eq. Eb with eta}
\end{equation}
where $E^{2}:=\Delta^{2}-E_{b}^{2}$.  In the particle-hole symmetry and non-magnetic sector $\eta=0$ the bound-state equation is reduced to 

\begin{equation}
E_{b}\mp\Delta\cos\frac{\phi}{2}+\left(E_{b}-\alpha(U)E_{b}\pm \beta\left(U,\phi\right)\Delta\right)\frac{\sqrt{\Delta^{2}-E_{b}^{2}}}{2\Gamma}=0.
\label{eq. Eb}
\end{equation}

Let us briefly discuss the phase dependence of the induced superconductivity.  We recall that by definition $\beta$ is $\Delta_{d}/\Delta$ to first order. Usually in non-self-consistent solutions, such as Matsumoto in Ref. \onlinecite{matsumoto2000}, one `chooses'  the phase dependence of $\Delta_{d}$ such that a smooth phase dependence of the localized excited states is achieved. However, the exact phase dependence of $\Delta_{d}$ can be determined by solving self-consistency equations.  In the high transmission case that we study here, we assume that the phase of the induced gap at the impurity site is governed by the two reservoirs and therefore one can choose $\Delta_{d}$ to have a $\cos(\phi/2)$ dependence on the phase. This allows the $\beta$ function to save this phase relation up to the second-order . On the other hand, the induced superconducting correlations at the impurity site are expected to drop through the repulsive Colulomb interaction therefore one can expect $\Delta_{d}$ to decrease with the increase of Coulomb repulsion\cite{{bauer},{matsumoto2000}}.

From Eq. (\ref{eq. Eb}) one can easily deduce that the only allowed bound-state at $\phi=\pi$ is $E_{b}=0$. Therefore, we consider the ansatz $E_{b}=\Delta f(U, \phi) \cos \phi/2$ as the bound-state in a high transmission channel. Substituting this ansatz into Eq. (\ref{eq. Eb}) the function $f(U,\phi)$ has to be the solution of 
\begin{eqnarray} \nonumber
&& \left(\cos^{2} \frac{\phi}{2} (1-a)^{2}\right) f^{4} \pm \left(2b \cos^{2} \frac{\phi}{2} (1-a)\right) f^{3} \\ \nonumber && +\left(\gamma^{2}+b^{2}\cos^{2} \frac{\phi}{2} - (1-a)^{2}\right)f^{2} \\ && \pm \left(-2 \gamma^{2}- 2 b + 2 a b \right) f + \gamma^{2}-b^{2} = 0.
\label{eq. Eb 4thdegree}
\end{eqnarray}
where  $\gamma:=2\Gamma/\Delta$.

\begin{figure*}[!ht]
 \includegraphics[width=5.2cm]{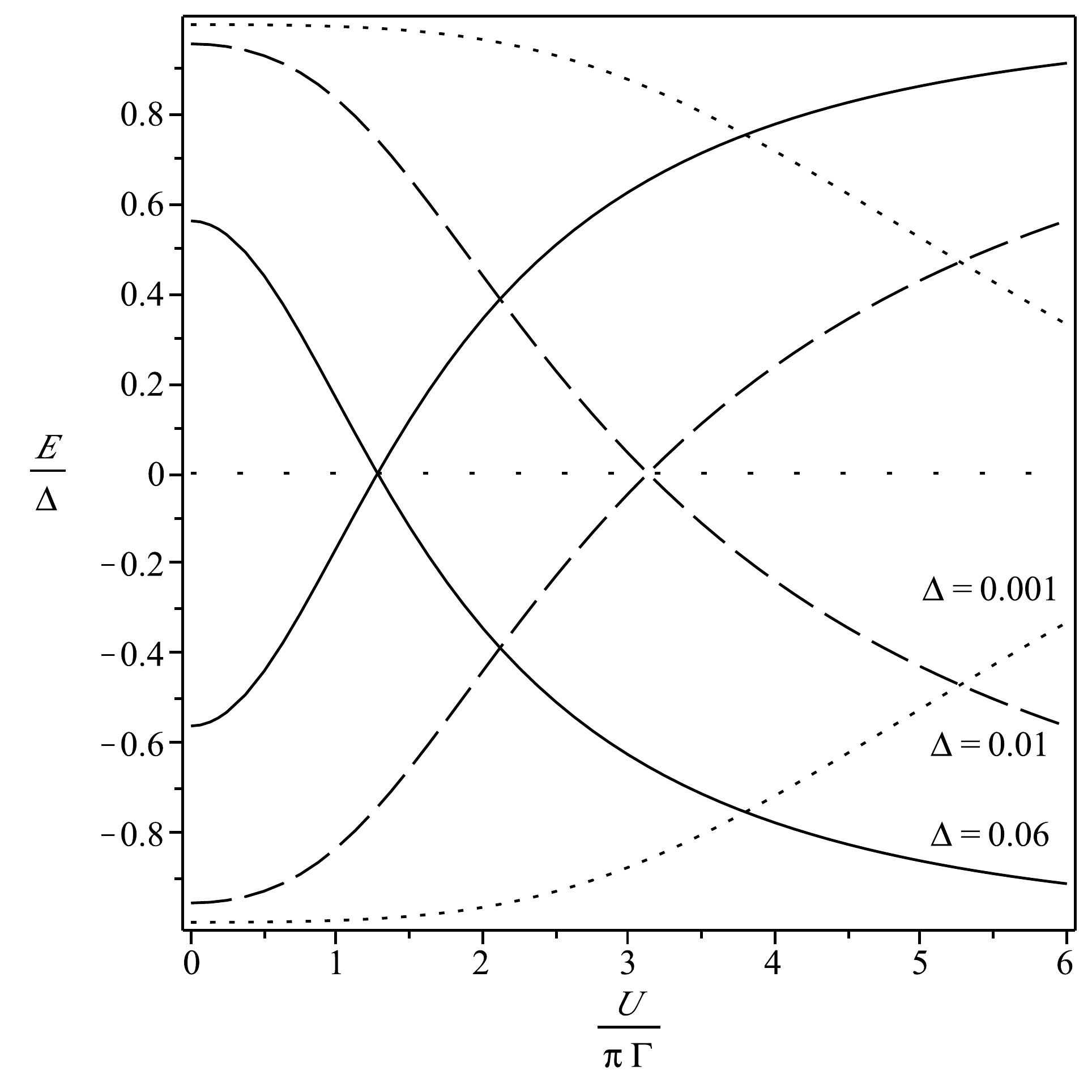}{(a)}\includegraphics[width=5.2cm]{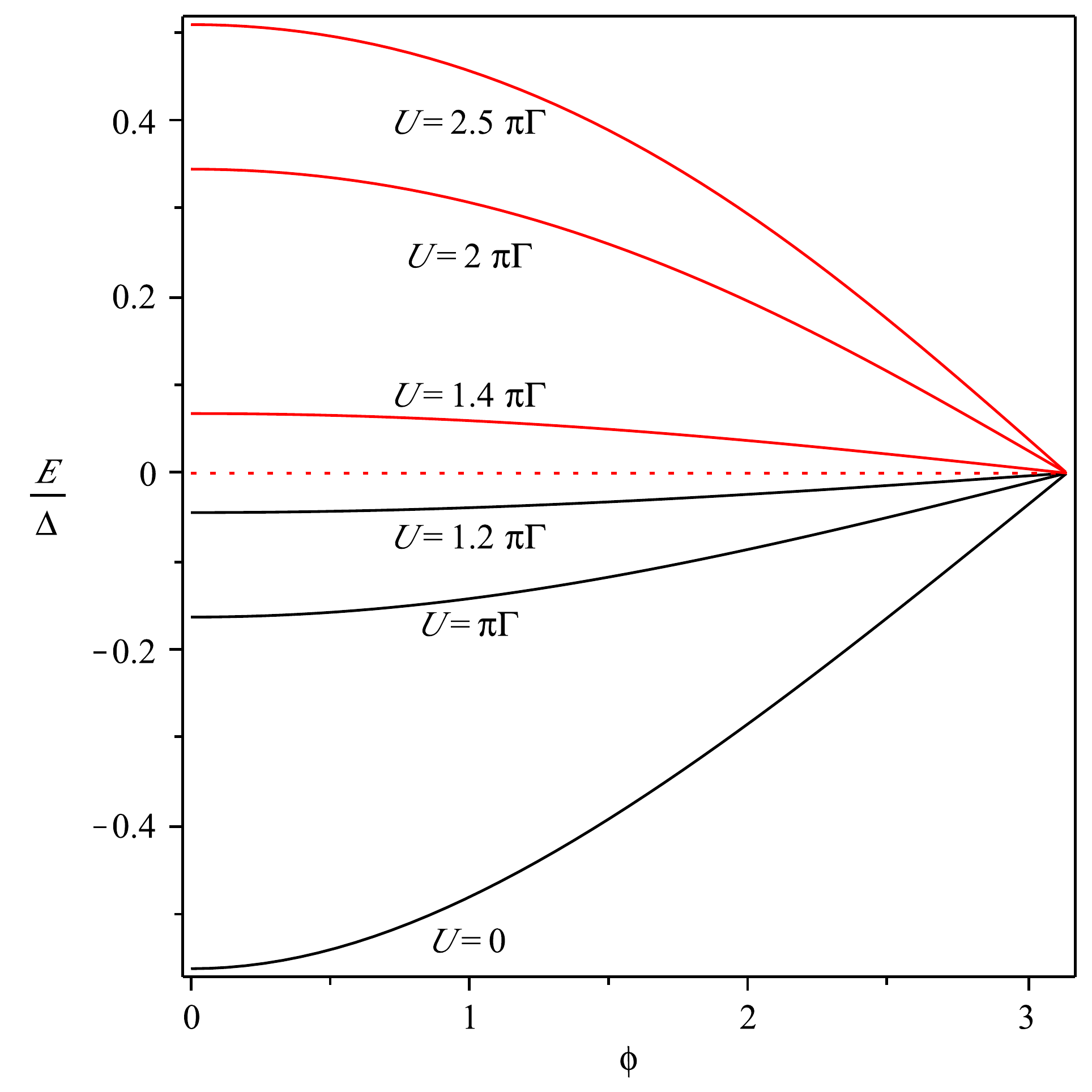}{(b)}
  \includegraphics[width=6cm]{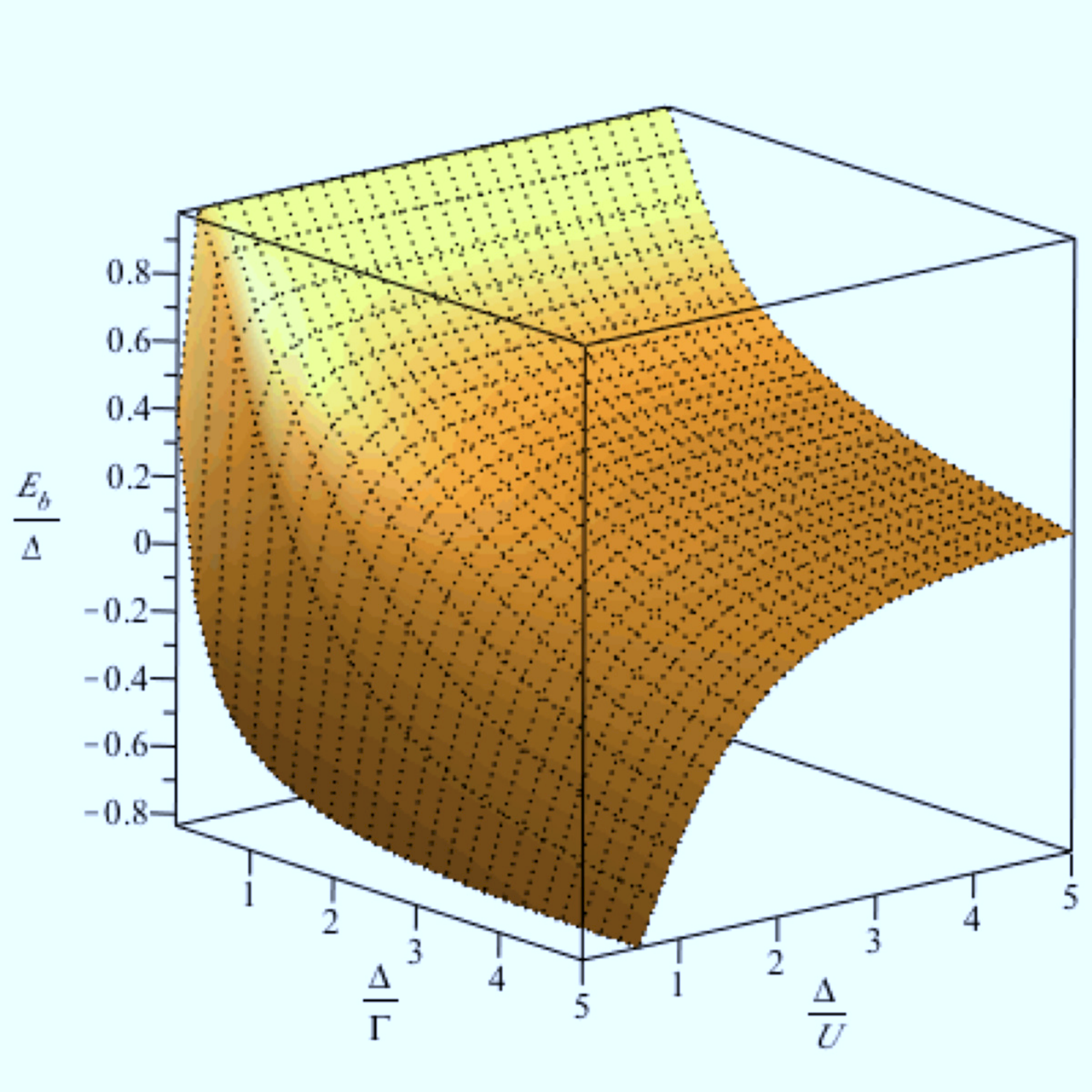}{(c)}  
 \caption{(Color online) Bound states at (a) zero phase limit $\phi=0$ due to the localized magnetic impurity inside a Josephson junction between two superconducting reservoirs, where $\Delta$ on the solid line is 0.06, dashed line 0.01, and dotted line 0.001; (b) in variable phase $0<\phi<\pi$ for different $\frac{U}{\pi\Gamma}$, where $\Delta=0.06$. In both plots $\pi\Gamma = 0.01$. (c) The bound-state in terms of $\Delta/\Gamma$ and $\Delta/U$ for the zero-phase case ($\phi=0$).}
\label{Fig. Eb} 
\end{figure*}

Fig. (\ref{Fig. Eb}a) presents the sensitivity of bound states to the Coulomb interaction $U$, extracted from the solution of Eq. (\ref{eq. Eb 4thdegree}). In the domain of second-order  perturbation theory  these states coincide with the numerical results taken recently from NRG method\cite{bauer}. Here we use those parameters used in Fig. (3) of Ref. [\onlinecite{bauer}] ---with a remark that the total hybridization strength  with both reservoirs $\Gamma_{L}+\Gamma_{R}=2\Gamma$ in our work is equivalent to the hybridization strength $\Gamma$ in Ref. [\onlinecite{bauer}]  where a single reservoir was considered.  

By the increase of the Coulomb repulsion the bound states first  approach the Fermi surface and, at the critical repulsion, overlap with the Fermi surface. Moreover, one can see in Fig. (\ref{Fig. Eb}a) the larger the gap is, the smaller the critical Coulomb repulsion becomes. Note  that this $\Delta$-dependence of the critical repulsion can be seen only after the second-order  is involved in the self-energy. This is the reason why this phenomenon cannot be seen in the first-order studies, neither in analytical \cite{matsumoto2000} nor in self-consistent first-order solutions.\cite{yoshiokaOhashi}

The most relevant regime for the parametrization is the large gap regime. We can study this regime by parameterizing our bound states in terms of $\Delta/2\Gamma$ and $\Delta/U$. The resulting bound states are plotted in Fig. (\ref{Fig. Eb}). 

The critical Coulomb repulsion can be found by setting bound states to zero in Eq. (\ref{eq. Eb}). This gives rise to the analytical formula for the critical repulsion
\begin{equation}
\beta\left(U_{c},\phi\right)=\gamma\cos\frac{\phi}{2}.
\label{eq. Uc}
\end{equation}
where $\gamma$ has been defined below Eq. (\ref{eq. Eb 4thdegree}).
 By substituting the off-diagonal
self-energy into Eq. (\ref{eq. Uc}) one can find the following equation whose solution gives rise to the critical Coulomb repulsion in terms of $\Delta$ and $\Gamma$. 

\beq
\left(\frac{\pi^{2}}{6}-1\right) \left(\frac{U_{c}}{\pi \Gamma}\right)^{2}\Delta\cos\frac{\phi}{2} + U_{c} \Delta_{d}=2\Gamma \cos\frac{\phi}{2}.
\label{eq. Uclarge}
\eeq

As
one can see in Eq. (\ref{eq. Uclarge}) for fixed $\Gamma$ and
increasing $\Delta$ the critical repulsion $U_{c}$ decreases as $\Delta^{-1/2}$. This behavior is in qualitative agreement with NRG results in that $U_{c}$ grows with falling $\Delta$ at roughly the same rate, albeit data are not sufficient to compare with the power law in detail \cite{bauer}.

As $U$ increases and crosses $U_{c}$ the spin orientation symmetry is broken and a magnetic moment forms, which indicates the magnetic impurity becomes less screened. This will introduce a nonzero $\eta$ in Eq. (\ref{eq. Eb with eta}). The bound states in the magnetic phase can only be found through the self-consistent analysis of the problem as the magnetic moment (say $\eta$) is determined at the same time as the bound states are. However in the non-magnetic phase we noticed because the bound states must be zero at $\phi=(\textup{odd\ number})\times \pi$ the bound states are of the form  $f(U,\phi) \cos(\phi/2)$, or in other words $f(U,\phi)\ (1-\tau \sin^{2}\phi/2)^{1/2}$ with $\tau=1$ for a ``ballistic'' channel. Beyond the non-magnetic phase one can expect the transmission rate $\tau$ to decrease with the rise of the magnetic moment in the junction and to depend on the spin polarization at the impurity site. It is natural to seek a solution in the magnetic phase, however, due to the inherited restriction of solutions in perturbative approaches in strong interactions we do not consider non-ballistic transmission in this paper.

%%%%%%%%%%%%%%%%%%%%%%%%%%%%%%%%%%%%%%%%%%%%%%%%%%%%%%%%%%%%%
%%%%%%%%%%%%%%% 2ND ORD:  .... SPECTRAL WEIGHTS  %%%%%%%%%%%
%%%%%%%%%%%%%%%%%%%%%%%%%%%%%%%%%%%%%%%%%%%%%%%%%%%%%%%%%%%%%

\subsection{Spectral weights}

The weight of the bound states is determined by evaluating the residue of impurity
Green's function at the bound states.  The normal (diagonal)
and anomalous (off-diagonal) weights are denoted by $W$ and $W_{\Delta}$. Having
defined $z(U)^{-1}=1-\alpha(U)$, one can find the weights from Eq. (\ref{eq. G}) as

\begin{widetext}
\begin{eqnarray}
W(\pm E_{b},U,\phi) & = & \frac{1}{2}\ \frac{\frac{E}{2\Gamma}+z\pm\frac{\eta zE}{2\Gamma E_{b}}}{1+\frac{E}{2\Gamma z}+\frac{E_{b}^{2}}{E^{2}}+\frac{\Delta^{2}}{E^{2}}z\beta(\phi,U)\cos\frac{\phi}{2}+\frac{2\Gamma\Delta^{2}}{E^{3}}z\sin^{2}\frac{\phi}{2}}
\label{eq. w}\\
W_{\Delta}(\pm E_{b},U,\phi) & = & \pm\frac{\Delta z}{2E_{b}}\ \frac{\cos\frac{\phi}{2}-\beta(U,\phi)\frac{E}{2\Gamma}}{1+\frac{E}{2\Gamma z}+\frac{E_{b}^{2}}{E^{2}}+\frac{\Delta^{2}}{E^{2}}z\beta(\phi,U)\cos\frac{\phi}{2}+\frac{2\Gamma\Delta^{2}}{E^{3}}z\sin^{2}\frac{\phi}{2}}
\label{eq. wD}
\end{eqnarray}
 \end{widetext}
 where $E=\sqrt{\Delta^{2}-E_{b}^{2}(U,\phi)}$. Note that in Eqs. (\ref{eq. w}) and (\ref{eq. wD}) $E$ and $E_{b}$ depend on both $U$ and $\phi$, and $z$ depends on $U$.

In the particle-hole symmetric case one can see that the normal weight is an even function of energy and the anomalous weight is an odd function, i.e., 
$W(E_{b},U,\phi)=W(-E_{b},U,\phi)$ and $W_{\Delta}(E_{b},U,\phi)=-W_{\Delta}(-E_{b},U,\phi)$. The asymmetric case needs a modification of the 
definite self-energies to become $\tilde{\alpha}$ and $\tilde{\beta}$, (e.g. see Eq.
[\ref{eq. modSE}) in Appendix B].

\begin{figure*}[!ht]
 % Requires \usepackage{graphicx}
 \includegraphics[width=7cm]{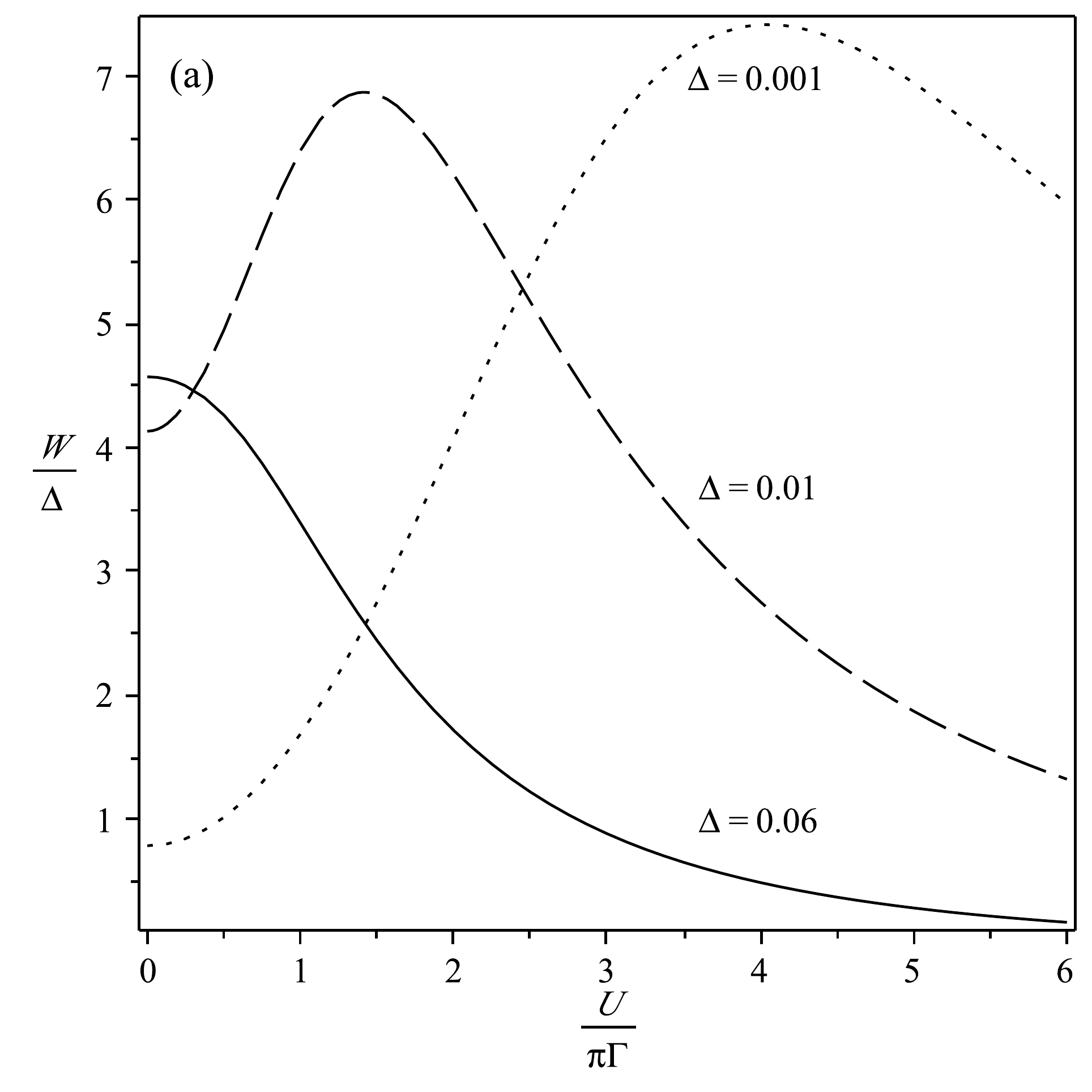} \includegraphics[width=7cm]{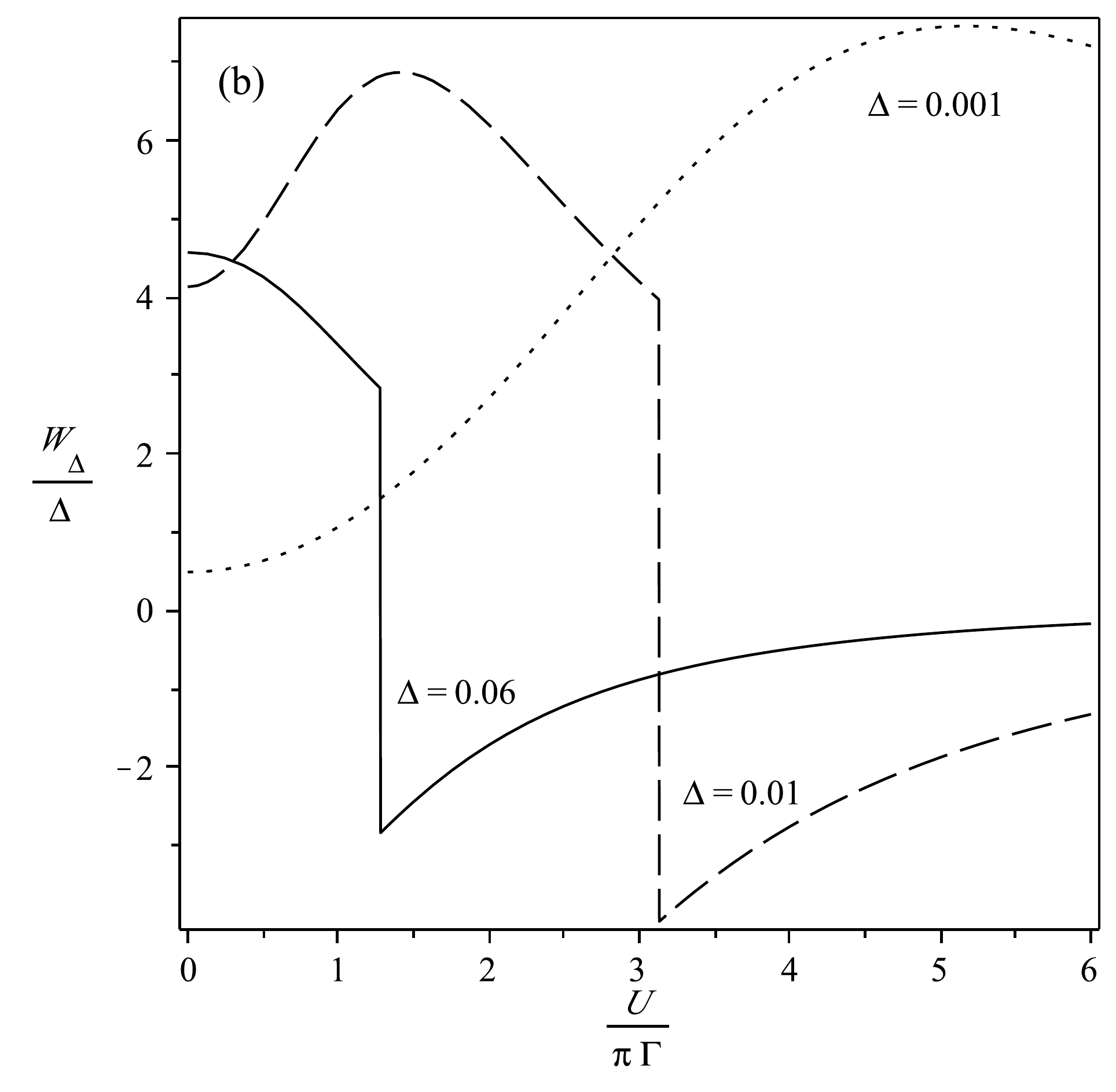} \\ \includegraphics[width=7cm]{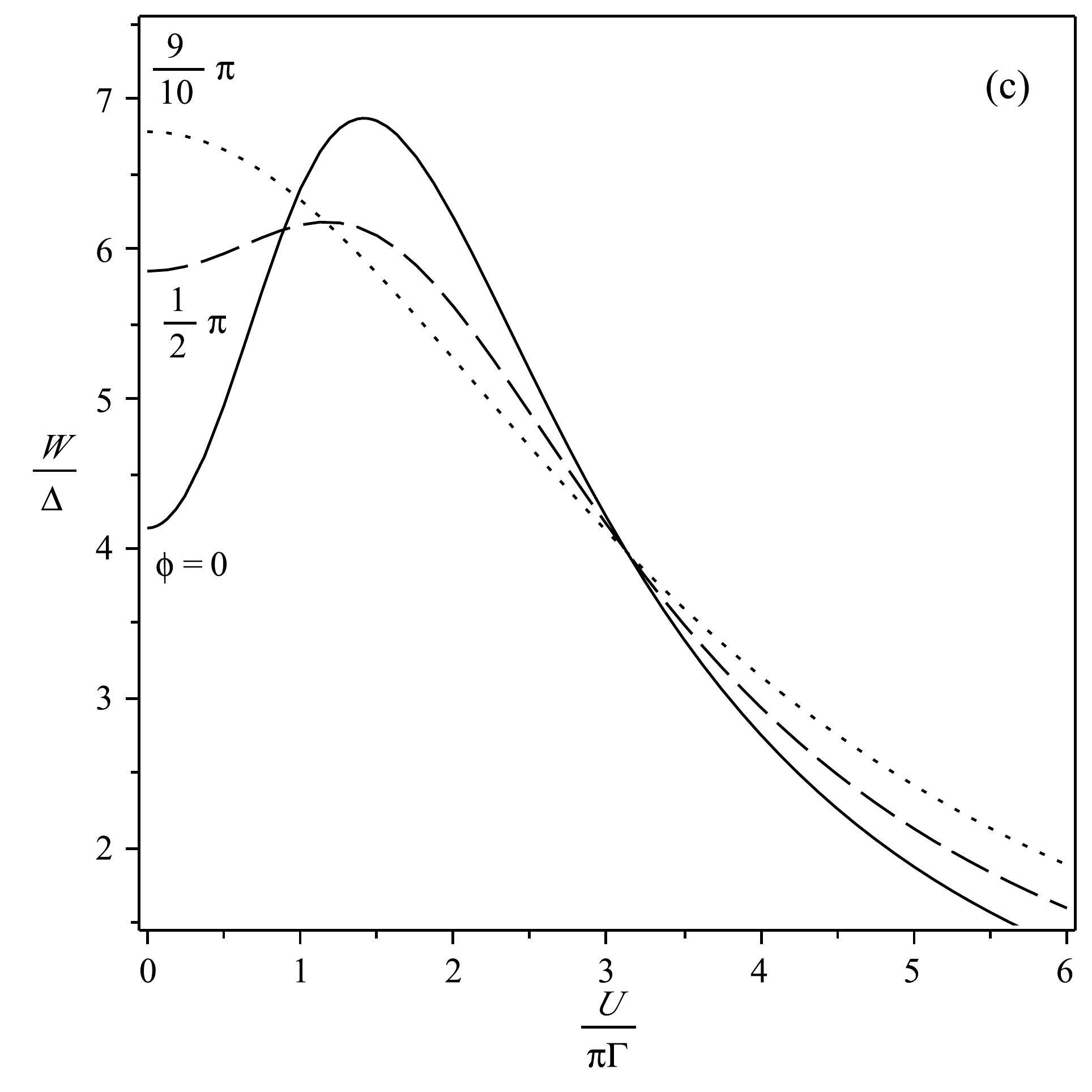}  \includegraphics[width=7cm]{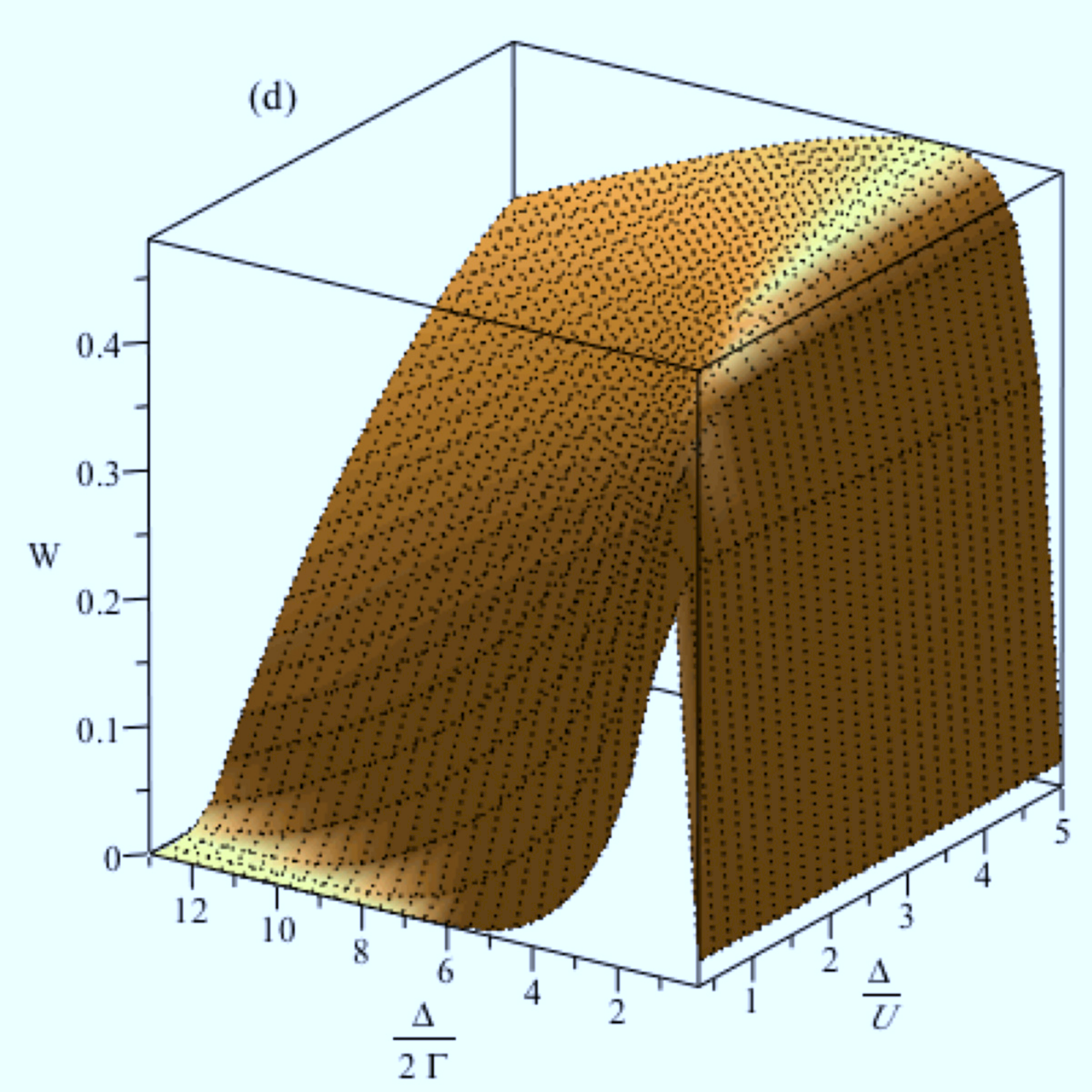} 
 \caption{ (Color online) (a) The spectral diagonal weight $W(E_{b})$, and (b)the off-diagonal weight $W(E_{b})$, of the sub-states whose energies are given in Fig. (\ref{Fig. Eb}).   (c) The phase dependence of diagonal weight; $\Delta=0.01$. {\bf d}) The normal weight as a function of $\delta=\frac{\Delta}{U}$ and $\theta=\frac{\Delta}{2 \pi \Gamma}$}
\label{Fig. weights} 
\end{figure*}

The diagonal and off-diagonal weights associated with those bound states whose energies were already given in Fig. (\ref{Fig. Eb}(a) are indicated in Fig. (\ref{Fig. weights}(a) and \ref{Fig. weights}(b), respectively.  The diagonal spectral weight for $\Delta=0.06$ and 0.01 are in good agreement with the corresponding NRG results;\cite{bauer} however, the criticality of $\Delta=0.001$ case, as indicated by NRG, is $\sim 5 \pi \Gamma$ which is beyond the validity of perturbation theory.

The off-diagonal weight $W_{\Delta}$ changes its sign at $U_{c}$. Due to the proportionality of $U_{c}$ and $\Delta^{-1/2}$ (via Eq. (\ref{eq. Uclarge})), one can expect that in larger gaps this change of sign is shifted to weaker interactions.  In next section we will see that the supercurrent flow inside a junction depends on the off-diagonal weight and therefore one expects that with decreasing Kondo temperature to below $0.3\Delta$ the supercurrent reversal occurs.  

Moreover, as the gap becomes smaller the maximum weight is shifted to higher Coulomb repulsions. Namely when the ratio of $\gamma$ is about 1 the maximum weight is at zero interaction and decreases monotonically in stronger repulsion; however for larger $\gamma$ the maximum is pushed to somewhere between zero and the critical repulsion.

\begin{figure}[h]
 % Requires \usepackage{graphicx}
 \includegraphics[height=6.5cm]{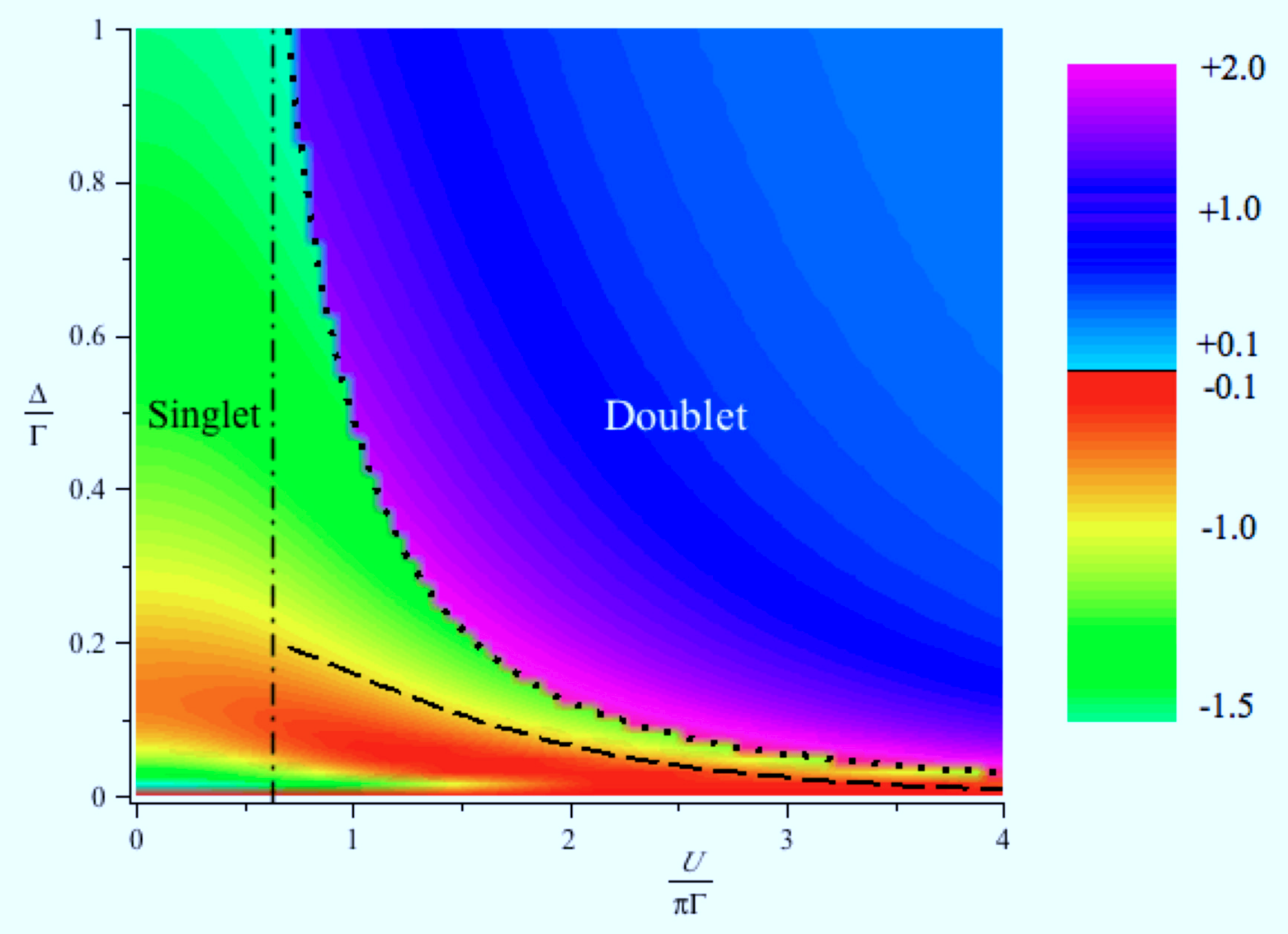} 
 \caption{(Color online) Phase diagram for singlet and doublet ground-states. The critical Coulomb
repulsion is denoted by $U_{c}$ to be compared with the doublet-singlet
boundary at $T_{K}/\Delta=0.3$. The dashed vertical line is the doublet-singlet
boundary at large $\Delta$ limit. The dash-dot line indicates the border between singlet and doublet states for very large gap superconductors. The off-diagonal weight $W_{\Delta}/\Delta$ is plotted in background colour. }
\label{fig. phase} 
\end{figure}

The competition between superconducting gap $\Delta$ and Kondo temperature $T_{K}$ becomes more clear in a phase diagram with the background color being the off-diagonal weight of the bound states. For the special case of zero phase this is depicted in Fig. (\ref{fig. phase}).

In this phase diagram the dotted line describes the phase boundary of Eq. (\ref{eq. Uclarge}), defined as the line where the bound states overlap the Fermi surface. This coincides with the boundary where the off-diagonal weight changes its sign.   The off-diagonal weight as discussed below Eq. (\ref{eq. wD}) should become negated as soon as the bound-state crosses the Fermi surface therefore it must either smoothly or suddenly change its sign about this point.  Let us consider a point very close to the Fermi surface  $E_{b}= \epsilon$ where $\beta \neq \gamma \cos(\phi/2)$.  Expanding the bound states from Eq. (\ref{eq. Eb}) about this point one can find $\epsilon/\Delta = \mp (\gamma \cos(\phi/2) - \beta)/(\gamma + 1 -\alpha)$. Substituting this energy into the off-diagonal weight one finds the numerator is finite valued in the vicinity of the Fermi surface while the denominator because of the presence of $E_{b}$ becomes very tiny.  This indicates that the off-diagonal weight should jump discontinuously from positive to negative when
crossing the Fermi surface.

This can be seen from Eq. (\ref{eq. wD}) where $E_{b}=0$ both the numerator and the denominator approach to zero. The off-diagonal weights, the pair amplitude, measure the degree of induced superconductivity. They connect to Ginzburg-Landau theory by taking the limit of interaction constant $g_{\rm BCS}$ to zero, [i.e., $\Psi e^{i\phi}=g_{\rm BCS} \langle d_{\up} d_{\down} \rangle$ in a superconductor]. Sign change in the right side can be interpreted as the phase shift by $\pi$ in the pair amplitude.
%note: I have spent five years evangelizing people to use the pair-amplitude 
%language and it is still the best one
 This phase lapse is traditionally called a $0-\pi$ transition. In the next section we will discuss this $0-\pi$ transition through the supercurrent reversal flow.   The dash-dotted line is the boundary at which the ground state makes a singlet-doublet transition in the large-$\Delta$ limit, (i.e., $U/\Gamma=2$).  Singlet ground states [i.e., $(|\rangle + |\up \down\rangle)/\sqrt{2}$] are on the left side of this border and the right side contains doublet groun states (i.e., $\{|\up\rangle, |\down \rangle\}$).\cite{{roderoPi},{bauer}} The dashed line gives the transition governed by the Kondo temperature $T_{K}=0.182U\sqrt{8\Gamma/\pi U}\exp(-\pi U/8\Gamma)$. Numerical analysis\cite{{yoshiokaOhashi},{satori92},{sakai}} indicates  that $T_{K}=0.3 \Delta$ is approximately the phase boundary between  magnetic and non-magnetic phases. This line is almost in the narrow region between the $U_{c}$ border and the maximum off-diagonal weight. 

One can see in Fig. (\ref{fig. phase}) that the critical Coulomb
repulsion line approaches  the phase boundary line $T_{K}/\Delta=0.3$
in the large $U$. This indicates that in large $U$ limit the crossover
of bound states occurs in the vicinity of singlet-doublet transition

%%%%%%%%%%%%%%%%%%%%%%%%%%%%%%%%%%%%%%%%%%%%%%%%%%%%%%%%%%%%%
%%%%%%%%%%%%%%% S U P E R C U R R E N T %%%%%%%%%%%%%%%%%%%%%%%%%%%%%
%%%%%%%%%%%%%%%%%%%%%%%%%%%%%%%%%%%%%%%%%%%%%%%%%%%%%%%%%%%%%

\section{Supercurrent}

%%%%%%%%%%%%%%%%%%%%%%%%%%%%%%%%%%%%%%%%%%%%%%%%%%%%%%%%%%%%%
%%%%%%%%%%%%%%% SUPERCURR:    .....  DEFINITION  %%%%%%%%%%%%%%%
%%%%%%%%%%%%%%%%%%%%%%%%%%%%%%%%%%%%%%%%%%%%%%%%%%%%%%%%%%%%%

\subsection{Definitions}

The Josephson current in a junction is derived from charge conservation
$\dot{n}_{d}+J_{L}-J_{R}=0$ where $n_{d}=n_{L}+n_{R}$ and $\dot{n}_{j}=[n_{j},H]/\imath\hbar$.
By substituting the Hamiltonian of Eq. (\ref{eq:eq H}) and defining
$\Delta e^{\imath\phi_{j}}\sim\sum_{k}\bra c_{k,\up}^{j}c_{-k,\down}^{j}\ket$, the current operator depends on impurity-reservoir Green's function $\mathcal{G}^{dj}$, i.e., $J_{j}=(2e t/ \hbar)\textup{Im}\intff d\omega\Tr(\mathcal{G}_{>}^{jd}(\omega)-\mathcal{G}_{>}^{dj}(\omega))$. Due to the Dyson equations the impurity-reservoir Green's function depends explicitly on the impurity ($\mathcal{G}^{d}$) and the reservoir ($g^{i}$) Green's functions, i.e.,  namely $\mathcal{G}_{r/a}^{dj}(\omega)=tg_{r/a}^{j}(\omega)\sigma_{z}\mathcal{G}_{r/a}^{d}(\omega)$. The Schwinger-Keldysh Green's functions are determined by the retarded
and advanced Green's functions $\mathcal{G}_{>}(\omega)=(1-f(\omega))(\mathcal{G}_{r}(\omega)-\mathcal{G}_{a}(\omega))$
and $\mathcal{G}_{<}(\omega)=-f(\omega)(\mathcal{G}_{r}(\omega)-\mathcal{G}_{a}(\omega))$
where, in thermal equilibrium, $f$ is the Fermi function at temperature $T$. Superconducting Green's function is known to be  
$(g^{j})_{11}=(g^{j})_{22}=\pi \rho \omega/E(\omega,\Delta_{j})$ and
$(g^{j})_{12}=(g^{j})_{21}^{*}=\pi \rho  \Delta_{j}e^{\imath\phi_{j}}/E(\omega,\Delta_{j})$, where $\rho$ is the density of states at the reservoirs. One can easily show at finite temperature that the current density
{between the impurity and the
$j$-th reservoir }becomes

\beq \label{eq. J} \langle J^{j}\rangle=\frac{8et^{2}}{\ensuremath{\hbar}} \textup{Im} \intff \left[ g^{j}(\omega)_{21}\ G^{d}(\omega)_{12}\right] f(\omega)d\omega.
\eeq

Without an impurity and identical reservoirs this current turns out to be sinusoidal with respect to the phase difference $\phi$. However, in the presence
of a magnetic impurity this sinusoidal current-phase
relation is deformed due to the presence of subgap states. Note that within the Hamiltonian model of 
Eq. (\ref{eq:eq H}) the Josephson current is conserved: $\bra J^{L}\ket=-\bra J^{R}\ket$.

The impurity Green's function describes the discrete bound states in the subgap region as well as the continuum spectrum above the gap. The bound states appear as poles of the Green's function. Accordingly, the Josephson current has two contributions: 1) from the continuum, $J_{c}^{j}$, which appears by tunneling into supergap
states at the impurity site, and 2) from bound-state current
that appears due to the tunneling of electrons into subgap states.
Some collective properties of both have been studied using NRG approach by Choi\emph{et al.}2004.\cite{choi} More recently,  Karrasch Oguri and Meden studied the total supercurrent by NRG and compared with the functional renormalization group (fRG) methods\cite{karrasch}. As a result the total supercurrent manifests a rather complicated $\phi$ dependence.  With the increase
of $U$ the current starts to be suppressed, until it starts to flow backward near the critical $U_{c}$.

Mean field theory can also predict the current reversal, however it lacks the Kondo
temperature as discussed in Sec. I. Let us now study the current flow in the second-order . We will see that the current flow can exhibit a controllable flow in the reverse direction in the strong-interaction limit. 

%%%%%%%%%%%%%%%%%%%%%%%%%%%%%%%%%%%%%%%%%%%%%%%%%%%%%%%%%%%%%
%%%%%%%%%%%%%%% SUPERCURR:  ....  SUBGAP CONTRI %%%%%%%%%%%%%%%%%%
%%%%%%%%%%%%%%%%%%%%%%%%%%%%%%%%%%%%%%%%%%%%%%%%%%%%%%%%%%%%%

\subsection{Subgap contribution to the supercurrent}

tunneling of Cooper pairs inside a junction between two superconducting reservoirs gives rise to the free energy of the junction $U(\phi)=-E_{J} \cos \phi$, where $E_{J}$ is the Josephson energy.  A nonzero value of $\phi$ will cause a supercurrent $I_{\textup{sc}}=(2e/\hbar) \partial U / \partial \phi= (2e E_{J}/\hbar) \sin \phi$ across the junction. 
In a tunnel junction without traps or time-reversal symmetry-breaking impurities, the 
Ambegaokar-Baratoff formula \cite{ambegbartt} predicts  $E_{J}=(\hbar \Delta / 4 e^{2 }R) \tanh (\Delta/2T)$, where $R$ is the normal state resistance of the junction and $T<T_c$.  At nonzero temperature due to the thermal activation of quasiparticles  there is also the normal current in addition to the supercurrent. However, at small temperature $T \ll 2E_{J}$ the effective resistivity of the tunnel junction becomes vanishingly small, thus the normal current becomes negligible and the dominant current will be the supercurrent, i.e., $I\ll I_{\textup{sc}}$.   This supercurrent is made of two parts: the continuous state and the subgap state parts: $I_{\textup{sc}}=I_{\textup{subgap}}+I_{\textup{cont.}}$  From the Ambegaokar and Baratoff formula one can expect that for temperatures much smaller than one half of the superconducting gap $T \ll \Delta/2$, the continuous
part becomes negligible, thus the subgap part becomes dominant. Here, we work out the subgap part of the supercurrent, which we expect to make the most crucial contribution.

The bound-state contribution sum up  over all bound states.  At zero temperature there is only the contribution of subgap states below the Fermi surface, therefore by substituting the off-diagonal spectral weight, instead of full Green's function, in Eq. (\ref{eq. J}) one finds the subgap contribution to the current as $\langle I_{j}\rangle=(16 e\Gamma \Delta/\hbar)  W_{\Delta}\sin(\phi/2)/ E $,
which by the use of Eq. (\ref{eq. wD}) turns into

\begin{widetext}
\begin{equation}
I_{\textup{subgap}}=\frac{4 e \Gamma \Delta^{2}}{\hbar}\frac{z}{E_{b}E}\frac{\sin\phi-\beta(U,\phi)\frac{E}{\Gamma}\sin\frac{\phi}{2}}{\frac{\Delta^{2}}{E^{2}}z\beta(\phi,U)\cos\frac{\phi}{2}+\frac{2\Gamma\Delta^{2}}{E^{3}}z\sin^{2}\frac{\phi}{2}+\frac{\Delta^{2}}{E^{2}}+\frac{E}{2\Gamma z}+1}\tanh\left(\frac{|E_{b}|}{2k_{B}T}\right)\label{eq. current}\end{equation}
\end{widetext}
where $E$ and $E_{b}$ depend on both $U$ and $\phi$, and $z$
depends on $U$. 

\begin{figure*}[!ht]
\includegraphics[width=8cm]{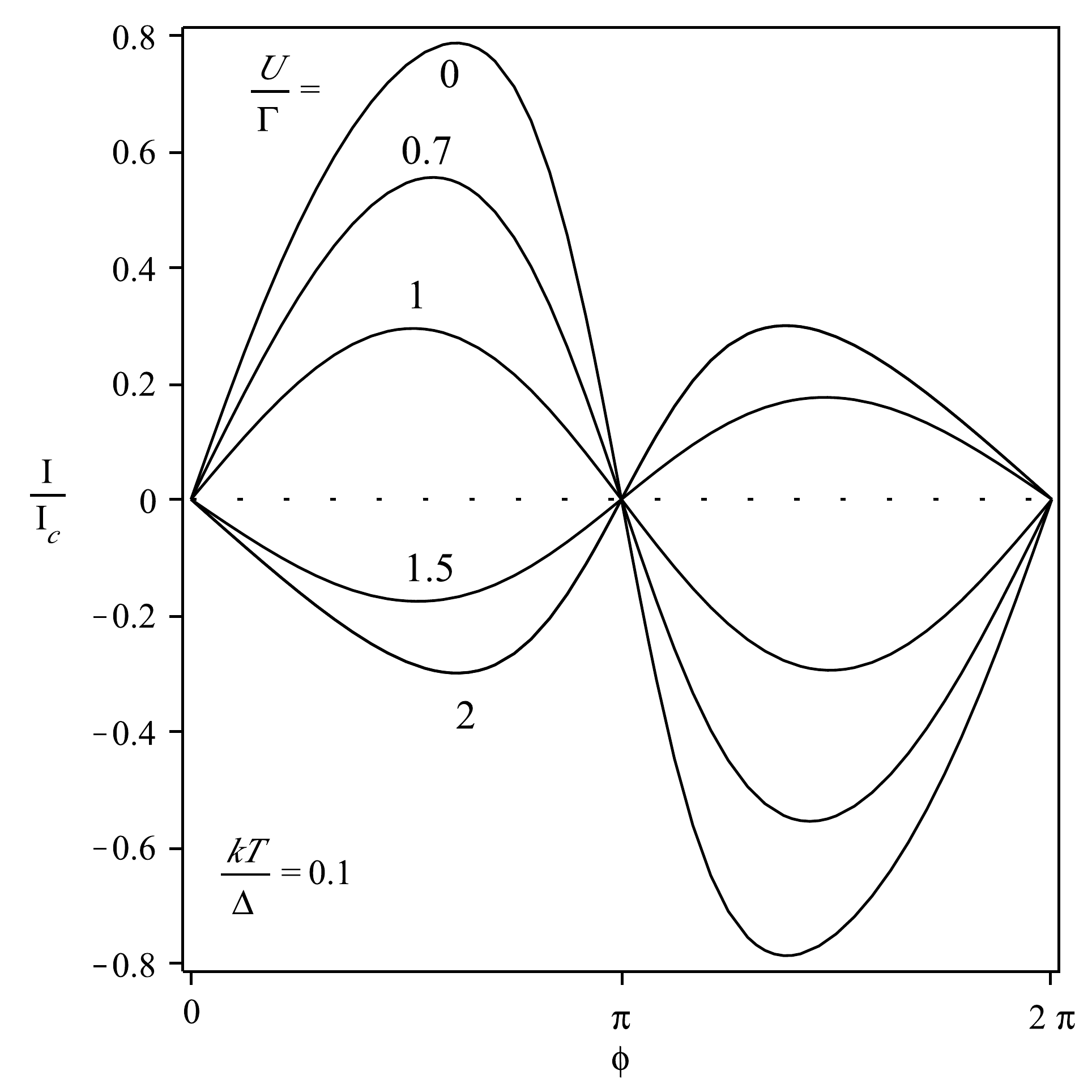} \includegraphics[width=8cm]{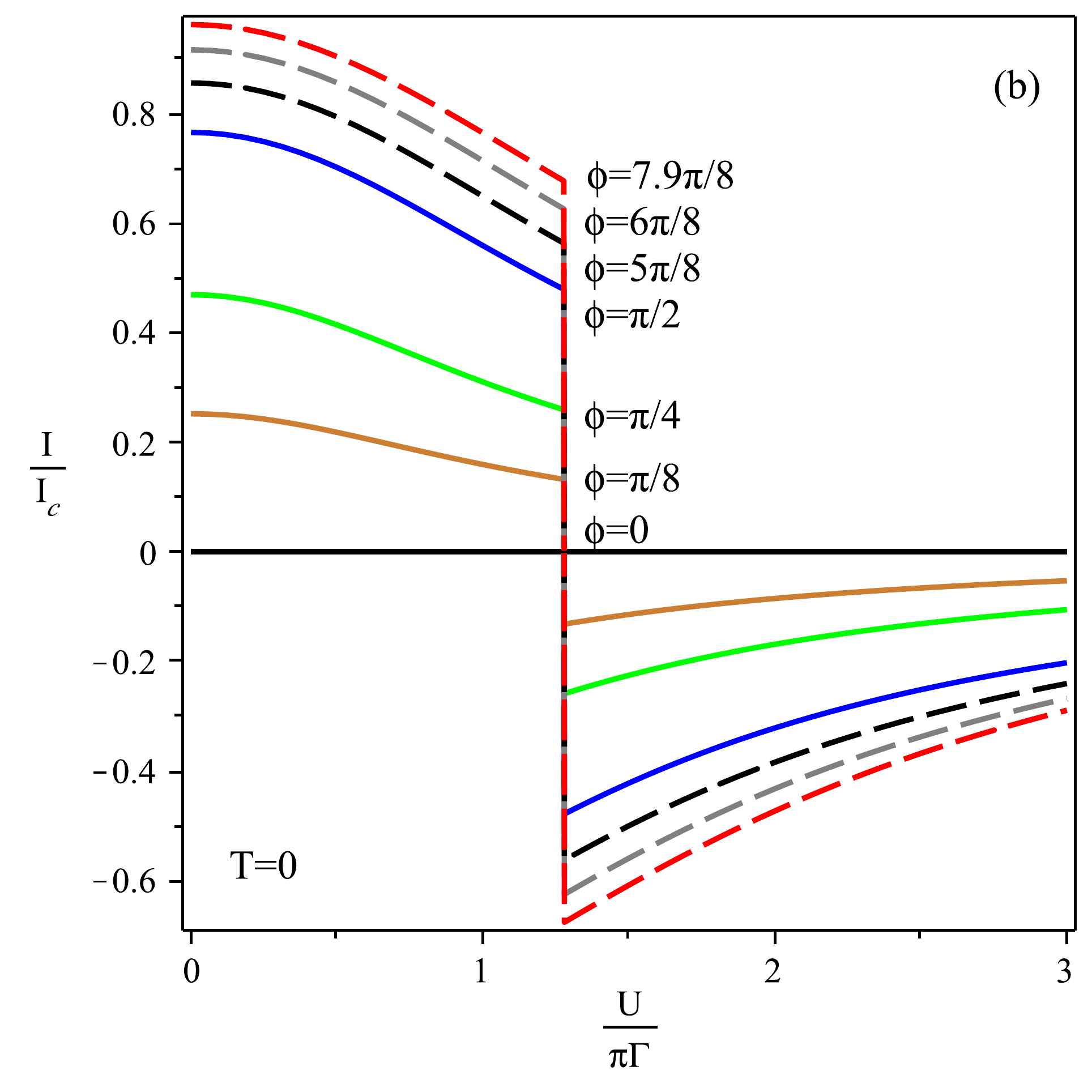} \caption{(Color online) Supercurrent variation with respect to (a) phase difference in different Coulomb repulsions at $kT=0.1\Delta$, (b) the Coulomb repulsion $U$ at different phases between 0 and $\pi$ at zero temperature; the reservoir-junction hybridization rate $\pi\Gamma=0.1$ and the gap $\Delta=0.01$. $I_{c}=\frac{2e \Delta}{\hbar}$.}
\label{Fig. IUphase} 
\end{figure*}

The subgap supercurrent dependence on $\phi$ at $T=0.1\Delta$ is shown in Fig. (\ref{Fig. IUphase}a) for different $U$. One can see the suppression by stronger repulsion and the change of sign at $U_{c}$.  Fig. (\ref{Fig. IUphase}b) shows the current reversal as a function of repulsion for different values of the phase at zero temperature. One can see that the extreme currents flow about the phase $\pi$. As indicated in Fig. (\ref{Fig. IUphase}a), by the temperature rise,  the maximum supercurrent, whose exact value depends on the $U$ and temperature, is shifted in the phase into between zero and $\pi$.  By increasing repulsion the extrema are shifted from nearby $\pi$ toward $\pi/2$ and after the crossover it is pushed back again toward $\pi$. This behavior has also be seen  in  NRG studies\cite{{choi},{karrasch}}.

Note that the functional renormalization group in Ref. [\onlinecite{karrasch}] shows that the supergap states may contribute to the $\phi$ dependence of critical repulsion $U_{c}$. As a result  the supercurrent reversal depend on both $U$ and $\phi$.  In other words, the criticality occurs for a particular repulsion at a particular phase. Let us emphasize  that here we restricted our analysis to the subgap states contributions and therefore our analysis lacks this peculiar phase dependence.   Nevertheless, what practically can be measured is not the exact supercurrent at a particular phase but instead the average of supercurrent at different phases.

%%%%%%%%%%%%%%%%%%%%%%%%%%%%%%%%%%%%%%%%%%%%%%%%%%%%%%%%%%%%%
%%%%%%%%%%%%%%% D E C O H E R E N C E %%%%%%%%%%%%%%%%%%%%%%%
%%%%%%%%%%%%%%%%%%%%%%%%%%%%%%%%%%%%%%%%%%%%%%%%%%%%%%%%%%%%%

\section{Current blocking decoherence}

Spectroscopy experiments on phase and flux qubits spectroscopically reveal
the presence of a few microwave resonators.\cite{{simmonds},{plourde}} These microresonators behave as spurious two-level systems inside the junction, whose coupling to the qubit produces reduced measurement fidelity and decoherence.\cite{cooper}. They can be intrinsically very coherent. \cite{Zagoskin06,Grabovskij11,Shalibo10} Recently, the trapping scenario for the dipole-dipole interaction between these states and the conduction electrons was studied by Constantin \emph{et al.}\cite{Constantin07}.  Using a different method, deSousa \emph{et.al.}\ cite{deSousa} considered the noise spectrum due to the occupation of a single localized magnetic impurity hybridized with electrons in a superconductor in the non-interacting regime. Their study shows the presence of some resonances in low temperature that could serve as candidates for the observed microresonators. However, the Kondo temperature in the non-interacting limit vanishes and their study does not include the competition between the Kondo temperature and the superconducting gap. On the other hand Faoro and Ioffe found that the underlying $1/f$  noise spectrum behaves correctly with the temperature [i.e., $S(\omega)\sim T^{2}$] only if the subgap bound states can be formed in the vicinity of the Fermi surface, which is not the case in the non-interacting Andreev bound states. As it was explained earlier in the paper the subgap bound states come closer to the Fermi surface if the Coulomb repulsion is taken into account. Therefore, Faoro and Ioffe concluded that the underlying $1/f$  noise must originate in interacting electrons.  Here, we study how the spectrum of additional microresonators may be modified in the presence of interaction.\cite{{Faoro05},{Faoro06},{Faoro07}} We will see the range of second-order  perturbation theory is sufficient to observe the emergence of novel
effects in the noise spectrum as the consequence of competition between superconducting gap and the Kondo temperature. We should note that a scenario based on local dipoles instead of traps has also been brought forward \cite{Constantin}.

In general the current noise spectrum is defined as $S=\int_{-\infty}^{\infty} dt \langle [\delta I(t),\delta I(0)  ]_{+}\rangle$. One can consider a constant hybridization coupling between the reservoirs and the junction and substitute the critical-current into this definition. The outcome from each magnetic impurity is known to be a
Lorentzian noise spectrum, and a combination of several
such spectra leads to $1/f$  noise. 

For the purpose of understanding the nature of a few microwave resonances on top of this noise spectrum one can assume  that the magnetic impurity partially block conduction whenever it captures
electrons from a reservoir. The model proposed by deSousa \emph{et.al.}\ cite{deSousa}  assumes that the channel average matrix element for electron
tunneling from one lead to the other depends on $n$
according to $t\sim t^{(0)}+t^{(1)}\hat{n}$. From Eq. (\ref{eq. J}) this can vary the supercurrent as $I_{c}=I_{c}(1+\hat{n} |t^{(1)}|/|t^{(0)}|)$. This modifies the current noise spectrum by the variation of number of electrons at the impurity site $\hat{n}$ such that $S_{\textup{total}}(\omega)=S_{I}(\omega)+(\delta I_{c})^{2} S_{n}(\omega)$. We shall study the contribution of the second term in the presence of Coulomb repulsion. This noise for $\omega>0$ is

\begin{widetext}
\begin{eqnarray}\nonumber
S_{n}(\omega)=\hbar \sum_{\sigma,\sigma'}\int_{-\infty}^{\infty} d\epsilon &&\left[\overline{\Gamma}_{\sigma\sigma';\sigma'\sigma}(\epsilon,\epsilon-\omega;\epsilon-\omega,\epsilon)\mathcal{A}_{\sigma, \sigma'}(\epsilon)\mathcal{A}_{\sigma, \sigma'}(\epsilon-\omega)\right.\\  
&& \left. - \overline{\Gamma'}_{\sigma\sigma';\sigma'\sigma}(\epsilon,\epsilon-\omega;\epsilon-\omega,\epsilon)\mathcal{B}_{\sigma', \sigma}(\epsilon)\mathcal{B}_{\sigma', \sigma}(\epsilon-\omega) \right] [1-f(\epsilon)]f(\epsilon-\omega)\\ \nonumber
\label{eq. noise}
\end{eqnarray}
\end{widetext}
where $\overline{\Gamma}_{\sigma\sigma';\sigma'\sigma}(\epsilon_{1},\epsilon_{2};\epsilon_{3}, \epsilon_{4})$ and $\overline{\Gamma'}_{\sigma'\sigma;\sigma\sigma'}(\epsilon_{1},\epsilon_{2};\epsilon_{3}, \epsilon_{4})$ are the diagonal and off-diagonal vertex corrections, respectively. The diagonal (off-diagonal) density of states $\mathcal{A}$ ($\mathcal{B}$) consists of the normal (anomalous) weight $W$ ($W_{\Delta}$) at the subgap states and the continuous function $A$ ($B$) for the supergap states. An analytical formula for the continuous part of the density of states is, to the best of our knowledge, unknown. However, numerical analysis\cite{{bauer},{hechtPaper}} indicate that with the increase of $U$ the continuous density of states above the gap is suppressed about the gap while two smooth maxima appear in high frequencies.

\begin{figure*}[!ht]
 % Requires \usepackage{graphicx}
 \includegraphics[width=11cm]{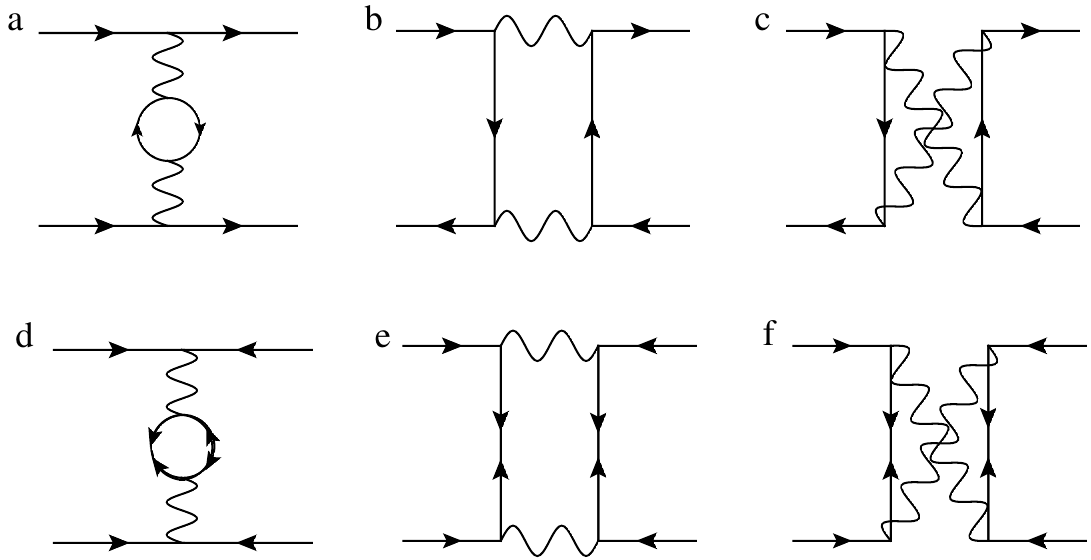} \caption{Vertex correction in the second-order .}
\label{Fig. vertexcorr} 
\end{figure*}

The noise spectrum has three major contributors in three frequency ranges (1) the subgap-subgap spectrum at low energy, (2) the subgap-continuous spectrum at moderate to high frequencies, and (3) the continuous-continuous spectrum in high frequencies. As qubits are operated at frequencies much below the gap, the first contribution is most relevant.
For the purpose of studying the microresonators that appear on top of the $1/f$  noise spectrum we compare the weights and frequency of peaks in the interacting noise power spectrum with that of non-interacting model\cite{deSousa}. The vertex correction diagrams   are shown in Fig. (\ref{Fig. vertexcorr}).  The diagrams  (a)--(c) represents the normal and diagrams (d)--(f) the anomalous contributions into the second-order  interaction of vertex correction. Analytical calculation of these diagrams shows that the diagrams (b) and (c) are exactly canceled out and so do the diagrams (e) and (f).  Moreover, the diagram (d) has a tiny contribution in the limit of $\Delta < \Gamma$ which is negligible.  One can calculate the vertex correction functions  $ \overline{\Gamma} \approx 1+(\Gamma^{2}\pi^{2}/8)(U/\pi \Gamma)^{2}$ and $\overline{\Gamma'}\approx 1$. The noise spectrum at low frequencies  can be separated in the following form: 

\begin{widetext}
\begin{eqnarray}\nonumber
S_{n}(\omega)/2\hbar= &&  \delta(\omega-2E_{b})\ \  2 W^{2} \Gamma [1-f(E_{b})]f(-E_{b})\\ \nonumber
&+& \theta(\omega-(\Delta-E_{b}))\ \ \{ W  [A(-E_{b}-\omega)+A(E_{b}+\omega))\overline{\Gamma}-2W_{\Delta} B(-E_{b}-\omega)\}[1-f(-E_{b})]f(-E_{b}-\omega) \\ 
&+&\theta(\omega-(\Delta+E_{b}))\ \ \{ W  [A(E_{b}-\omega)+A(-E_{b}+\omega))\overline{\Gamma}-2W_{\Delta} B(E_{b}-\omega)\}[1-f(E_{b})]f(E_{b}-\omega). \\ \nonumber
\end{eqnarray}
\end{widetext}

\begin{figure*}[!ht]
 % Requires \usepackage{graphicx}
\includegraphics[width=8.5cm]{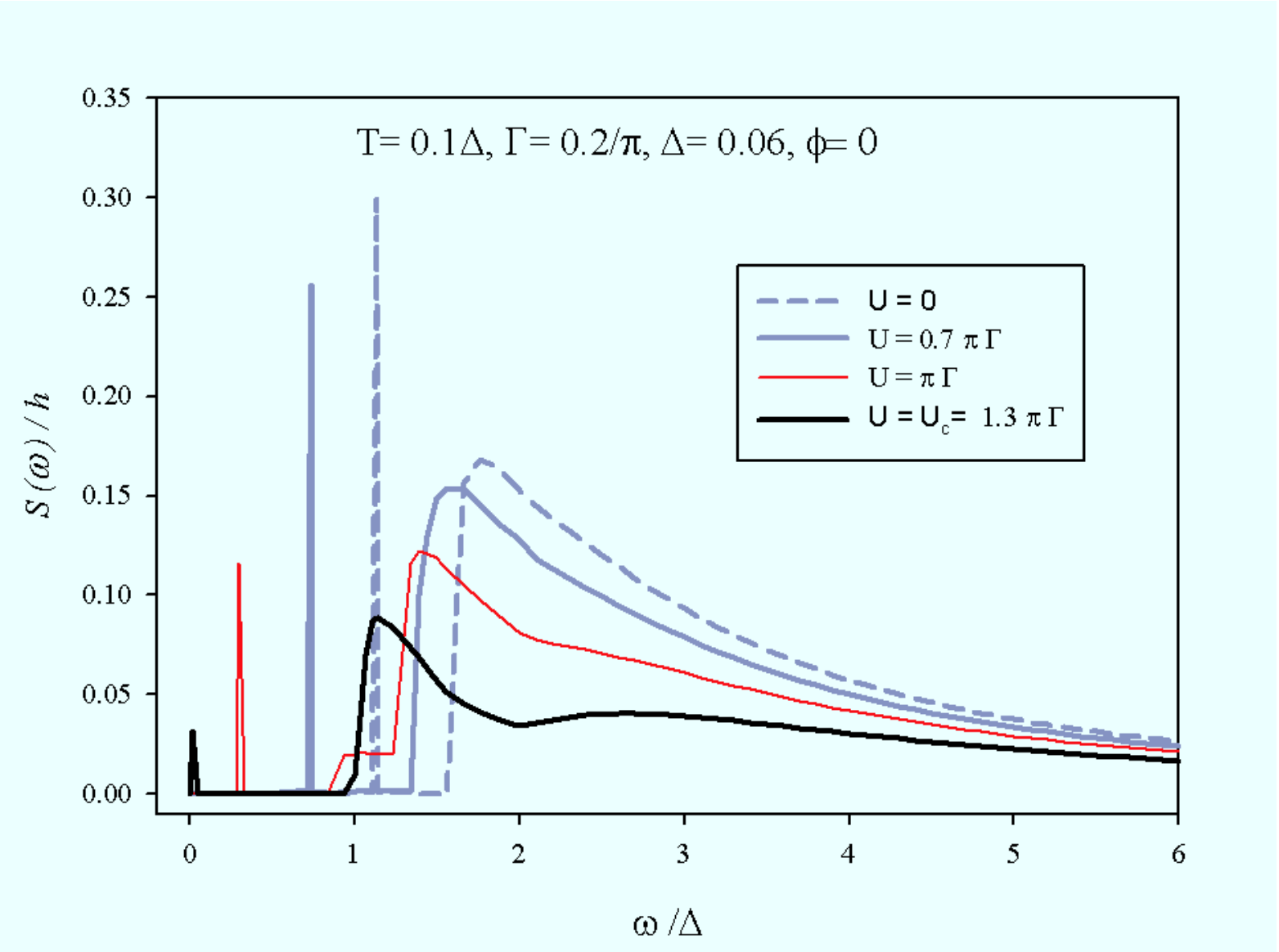} \includegraphics[width=8.5cm]{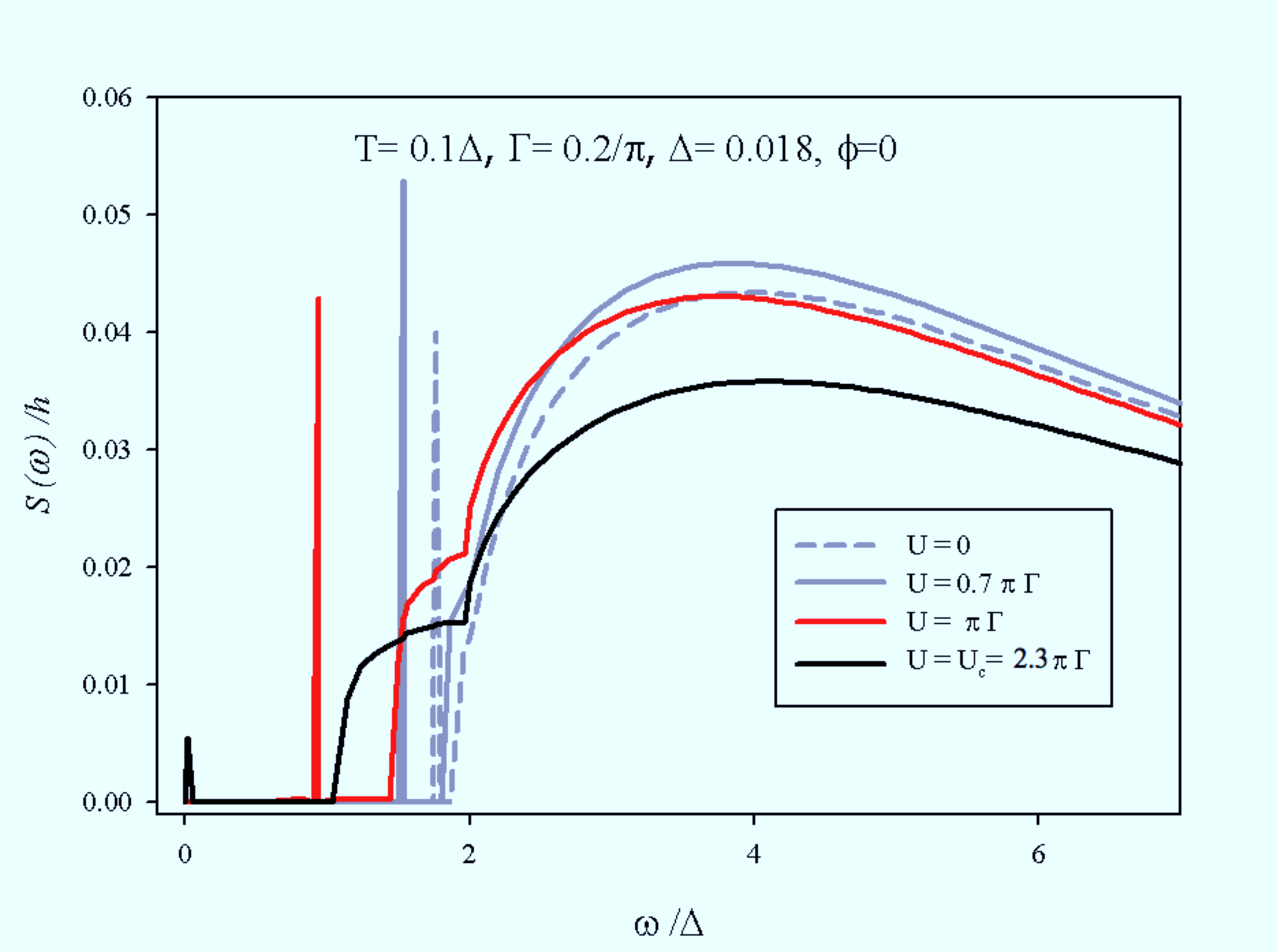}  \caption{(Color online) The charge blockade noise spectrum for various Coulomb repulsions as listed in the legend}
\label{Fig. SnWnomag} 
\end{figure*}

Note that with the increase of $U$ the subgap bound states approach  the Fermi surface. Within our focus of low frequencies two noise spectra are depicted in Fig. (\ref{Fig. SnWnomag}a,b).  

Fig. (\ref{Fig. SnWnomag}) presents  the charge blockade noise spectrum for various Coulomb repulsions. With the increase of Coulomb repulsion the spectral noise of subgap-subgap transition is affected in two ways: 1) the resonance frequency is lowered until it becomes zero at the critical repulsion. Stronger repulsions than the critical one are expected to push back the resonance into higher frequencies. 2) The noise spectral weight of the sub-gap resonance by the increase of repulsion decreases (except at small gaps where the noise may first increase and after achieving  a maximum at $U<U_{c}$ it suppress onward). As a typical example one can see that with the increase of Coulomb repulsion from zero to $U=0.2$ the weight of the resonance is suppressed by $60\%$.   We recall that the normal weight, shown previously in Fig. (\ref{Fig. weights}(a), in a large-gap superconductor ---i.e., $\Delta\sim \Gamma$---monotonically decreases with the increase of $U$. However, by increasing $U$ in smaller gap materials this weight may raise to a maximum  at $U_{\textup{max}}<U_{c}$ and then becomes suppressed in stronger interactions. This can be seen for two typical parametrization  of $\Delta/\Gamma=0.94$ and $\Delta/\Gamma = 0.28$ in Fig. (\ref{Fig. SnWnomag}(a) and Fig. (\ref{Fig. SnWnomag}(b), respectively.

\begin{figure*}[!ht]
 % Requires \usepackage{graphicx}
 \includegraphics[width=7.7cm]{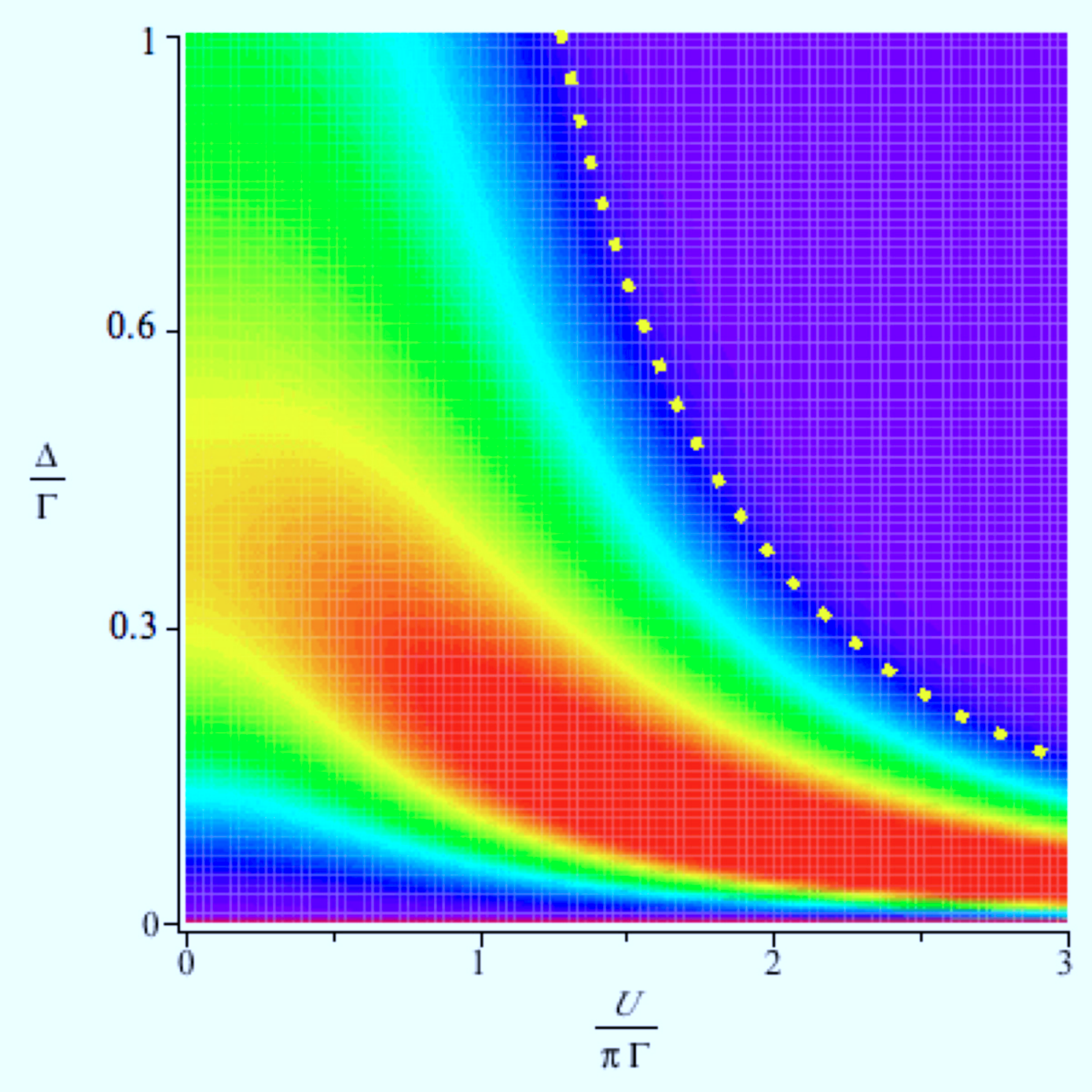}\includegraphics[width=10cm]{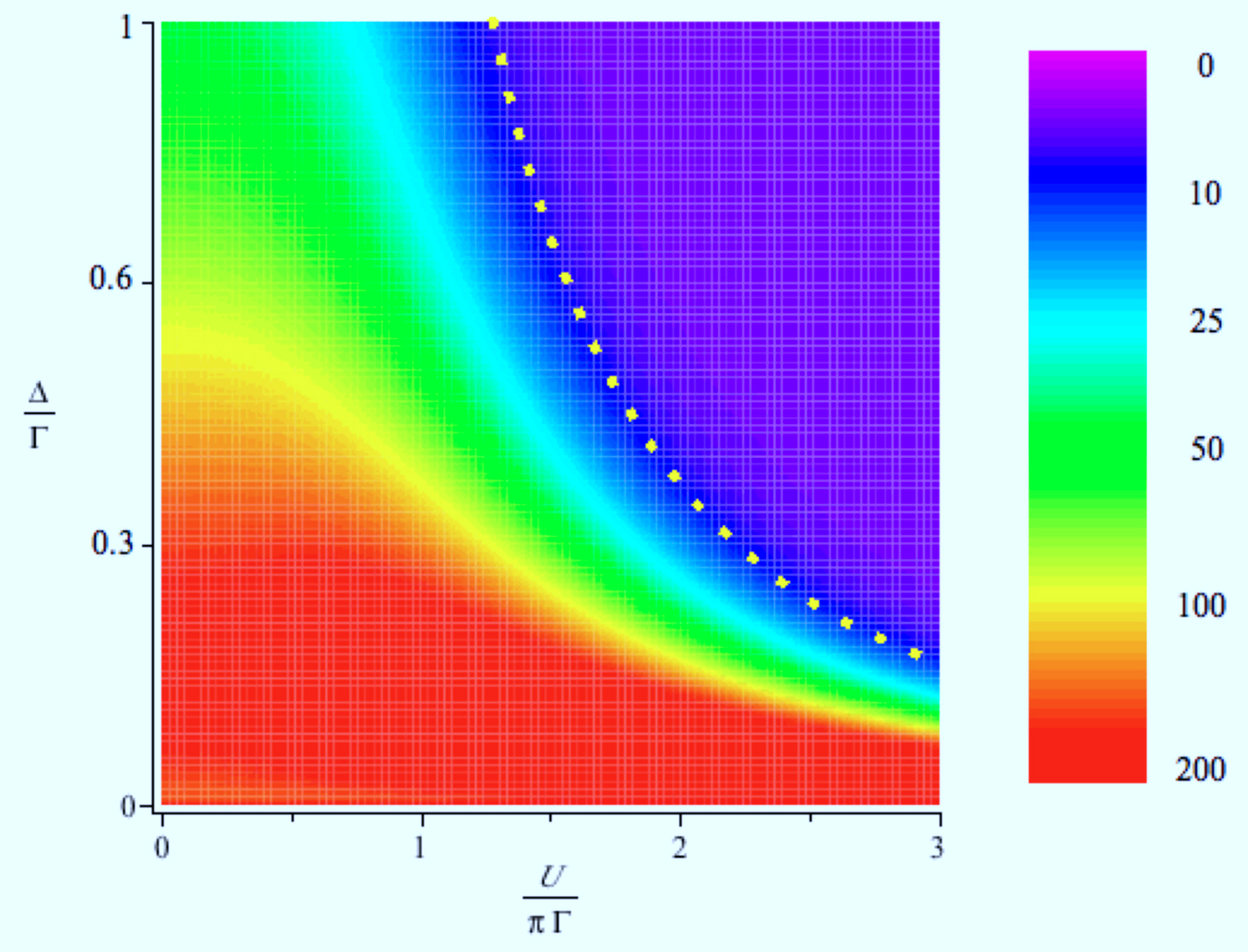}
 \caption{(Color online) The background colours indicate the spectral weight of the resonance noise as a function of ratio $\Delta/\Gamma$ and $U/\pi \Gamma$ for the phase difference $\phi=0$ (left) and $\phi=\pi/2$ (right). Dotted lines are the singlet-doublet boundary. }
\label{Fig. SwT} 
\end{figure*}

The resonance noise  is highly sensitive to the size of the superconducting gap, the hybridization strength, and the Coulomb repulsion between onsite electrons. The peculiar sensitivity of the spectral weight of this noise is plotted in Fig. (\ref{Fig. SwT}) for the two cases of $\phi=0$ and $\phi=\pi/2$. The dotted line indicate the singlet/doublet boundary.  The background color indicates $S(\omega)/\Delta^{2}$ with red the highest noise power that decreases into the lowest purple area through  yellow, green and  blue colours, respectively. In Fig. (\ref{Fig. SwT}) the left plot represents a zero-phase junction. The phase difference in the right plot is instead $\pi/2$. By comparing these two one can see in the latter case that the power of resonance noise monotonically suppresses with the increase of interaction in every parametrization of $\Delta/\Gamma$.

%%%%%%%%%%%%%%%%%%%%%%%%%%%%%%%%%%%%%%%%%%%%%%%%%%%%%%%%%%%%%
%%%%%%%%%%%%%%% C O N C L U S I O N %%%%%%%%%%%%%%%%%%%%%%%%%
%%%%%%%%%%%%%%%%%%%%%%%%%%%%%%%%%%%%%%%%%%%%%%%%%%%%%%%%%%%%%

\section{Conclusion}

We  presented an analytical formalism to study a magnetic impurity residing at the Josephson junction partially blocking the only ballistic channel of the junction in the limit of moderate Coulomb interactions and small superconducting gap.  We applied the second-order  perturbation theory in terms of the Coulomb repulsion and analytical formulate the subgap bound-state energy and their normal and anomalous weights. These results were shown to be in good agreement with recent results of NRG.  We expressed the supercurrent as a function of the Coulomb repulsion and noticed that in the strong-interaction limit the supercurrent may flow in the reverse direction compared to the noninteracting case.   We finally studied the noise generated due to  the partially blockade of conduction caused by the magnetic impurity. Depending on the fabrication of the superconducting gap, the hybridization strength and the Coulomb repulsion the resonant noise can be heavily suppressed from the superconducting material.  

Note that we do not have control on the magnetic impurity site. For instance the level position $\ed$, measured from the Fermi energy of the two leads is not  a parameter that is tuneable by an external gate voltage, as it is in the case of quantum dot. The interaction $U$ is not a controllable parameter either.  As a consequence we cannot think of the bound states as a qubit. These states play the role of  spurious qubits  which could strongly  couple to the desired qubit. However the microwave resonances they generate for the critical-current are highly sensitive to the interaction.  

 A quantum dot when coupled to two superconductors may behave as a magnetic impurity in the junction. One of the advantages of using a quantum dot is that the Coulomb repulsion is fully controllable in a quantum dot, thus one can test the suppression of the resonant noise spectrum.  In this paper,  we studied only the non-magnetic solution. Extending this study to cover the entire phase diagram, one needs to set up a self-consistent solution.

The consequences of this physics of superconducting qubits can be elucidated as follows. It is established that junctions contain trap states like ours. Similar to the noninteracting case \cite{deSousa}, depending on the trap energy, the junction resonators \cite{simmonds} can be anywhere within the superconducting gap, hence most of them won't ever resonate with the qubit whose energy splitting covers a band that is deep inside the gap and comparably narrow. Our work shows that Coulomb repulsion moves these resonators to lower frequencies hence toward the qubit frequency, while simultaneously reducing their spectral weight. With suitable distributions of both $\epsilon_d$ and $U$, we might expect $1/f $-type noise spectra in addition to the resonator noise. We also see that the $\pi$-junction mechanism, albeit intriguing, does not give a marked contribution to the noise.

A quantum dot as an artificial impurity in between superconducting reservoirs can be one option to test our predictions.   A quantum dot with an occupied electron breaks Cooper pairs to be screened and produce a Kondo singlet state. A variety of Kondo temperatures for the states with odd number of electron are accessible \cite{{vanDam:2006fj},{Buitelaar:2002gf}}, where negative critical-current is expected. Having access to both distinguishing the quantum dot states and their associated Kondo temperature (i.e., the effective Coulomb interactions) the phase decoherence of an initial state can reveal the underlying critical-current noise, which can be compared at different Kondo temperatures. The maximum decoherence is expected at a narrow zone of parameters. One can also think of applying a weak in-plane magnetic field to a phase qubit, or coupling it to a cavity in order to tune the Coulomb interaction in the junction and measuring the difference in the spectral density of the resonance peaks in the excitation frequency.

\begin{acknowledgements} 
MA should warmly thank Kosaku Yamada, Alfredo Levy-Yeyati, Irfan Siddiqi, and Britton Plourde for helpful discussions.  FW acknowledges an inspiring discussion with Volker Meden. This research was partly funded by the Office of the Director of National Intelligence (ODNI), Intelligence Advanced Research Projects Activity (IARPA), through ARO. 
\end{acknowledgements}

%%%%%%%%%%%%%%%%%%%%%%%%%%%%%%%%%%%%%%%%%%%%%%%%%%%%%%%%%%%%%
%%%%%%%%%%%%%%% A P P E N D I X %%%%%%%%%%%%%%%%%%%%%%%%%%%%%
%%%%%%%%%%%%%%%%%%%%%%%%%%%%%%%%%%%%%%%%%%%%%%%%%%%%%%%%%%%%%
\appendix
\renewcommand\thesection{ \Alph{section}}

%%%%%%%%%%%%%%%%%%%%%%%%%%%%%%%%%%%%%%%%%%%%%%%%%%%%%%%%%%%%%
%%%%%%%%%%%%%%% A %%%%%%%%%%%%%%%%%%%%%%%%%%%%%%%%%%%%%%%%%%%
%%%%%%%%%%%%%%%%%%%%%%%%%%%%%%%%%%%%%%%%%%%%%%%%%%%%%%%%%%%%%

\section{SELF-ENERGY}

Let us approximate the Green's function of a superconductor with a small
energy gap with simpler functions. The dominant behavior of the normal
Green's function in the small gap limits can be approximated by
a normal metal: 
\begin{eqnarray}\nonumber
G^{o}(\omega)&=&\includegraphics[width=3cm]{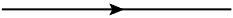}=\frac{1}{D(\omega)}\left[\omega\left(1+\frac{\Gamma}{E(\omega)}\right)+\epsilon_{d}\right] \\ && \approx\frac{1}{{\omega}+sgn(\omega){\Gamma}}
\label{eq. App G prop}
\end{eqnarray}
where $D(\omega)$ is the same as $F(\omega)$ denoted in the denominator of Eq. (\ref{eq. G}) for the noninteracting case ($U=0$). Note that in noninteracting particle-hole symmetry $\epsilon_{d} = 0$.   

The off-diagonal Green's function vanishes when the gap is set to zero;
however at small energies there is a finite contribution that can
be provided by taking the limit of $\omega\rightarrow0$ before considering
the small-gap limit. One can find this function as follows: 

\begin{eqnarray}\nonumber
F^{o}(\omega)&=&\includegraphics[width=3cm]{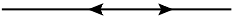}=\frac{1}{D(\omega)}\left[\frac{\Delta\Gamma\cos(\phi/2)}{E(\omega)}\right] \\ &&\approx \frac{\Delta \Gamma \cos\frac{\phi}{2}}{\omega \left(\omega+\Gamma \right)^{2}} 
\label{eq. App F prop}
\end{eqnarray}

Considering the Pauli exclusion principle the Coulomb repulsion takes effect between two electrons with opposite spins. In the limit of $\Delta \ll \Gamma$ the following diagrams take part into the self-energy matrix:

\
 \
\subsection{Diagram (a)}

\begin{figure}[h]
 \includegraphics[width=5cm]{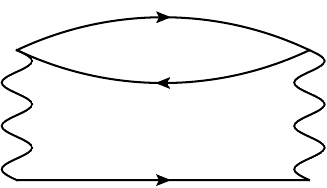} 
\end{figure}

\begin{figure}[h] \label{fig. integdomain}
 \includegraphics[width=8cm]{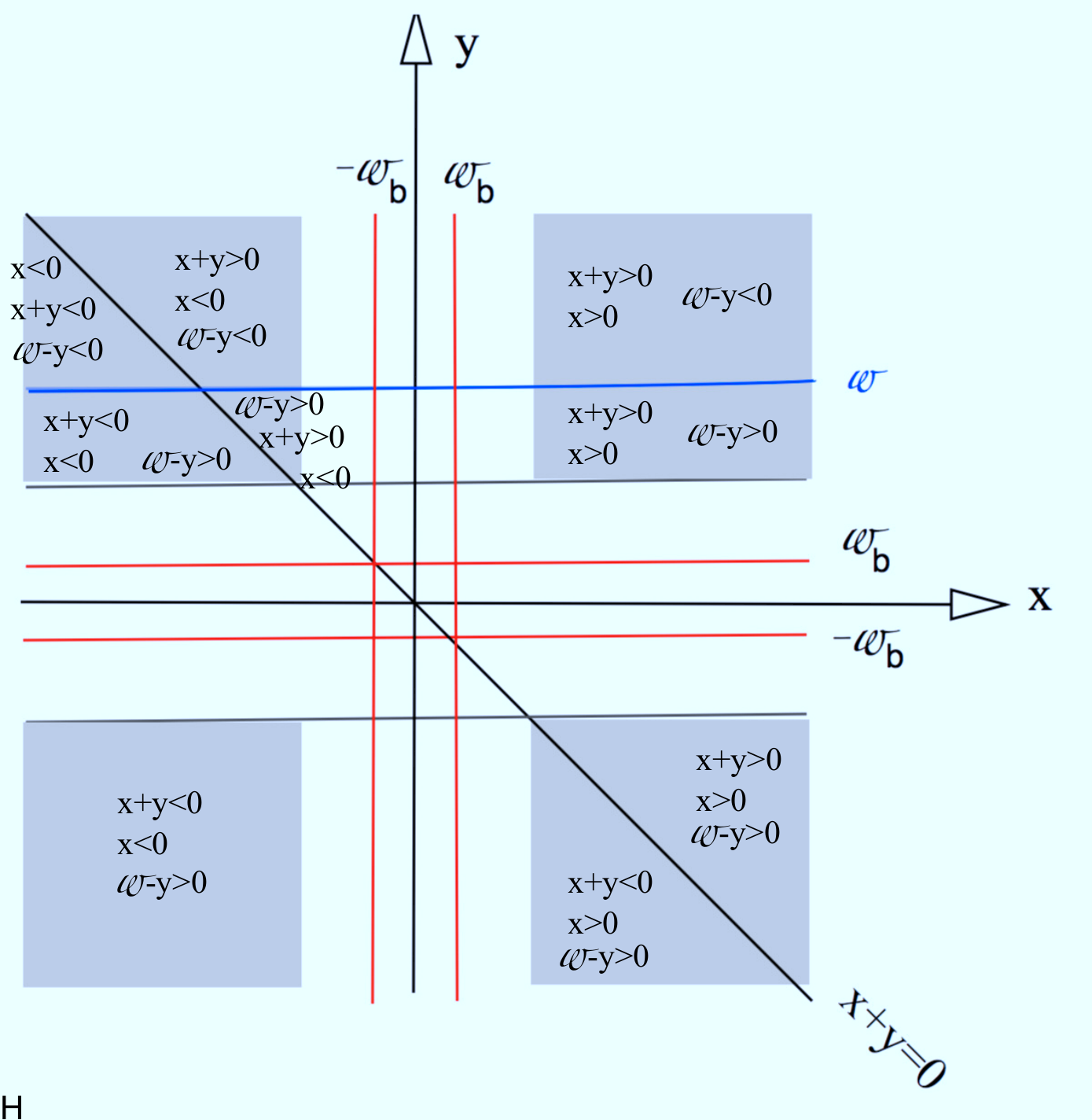} 
 \caption{The integration domain space.}
\end{figure}

From Eq.(2.8) in Yamada and Yoshida's paper\cite{{yamada},{yamada1},{yamada2}}, the send order term of $G(\tau,\tau')$ is given by

\begin{eqnarray*}
&&\frac{-U^{2}(-1)^{2}}{2} \{ G(\tau,1) G(1,2) G(\tau_{2},\tau')\\ &+& G(\tau,2) G(2,1) G(1,\tau')\} \ G(1,2)G(2,1),
\label{eq. GGGt}
\end{eqnarray*}
where the term within $\{\ \}$ is for up spin and $G(1,2)G(2,1)$ is for down spin. Let is calculate the first term in the Fourier space. Using Feynman diagram laws one can write 
\begin{widetext}
\begin{eqnarray*}
\Sigma(\omega) & = & U^{2}\int_{-\infty}^{\infty}\frac{dy}{2\pi}\, G^{o}(\omega-y)\:\int_{-\infty}^{\infty}\frac{dx}{2\pi}G^{o}(x+y)\, G^{o}(x)\\
 & = & U^{2}\left(W\left(-\omega_{b}\right)+\int_{-\infty}^{-\Delta}\frac{dy}{2\pi}\frac{1}{\omega-y+\Gamma}\right) \left(W\left(-\omega_{b}\right)^{2}+W\left(\omega_{b}\right)^{2}+\int_{-\infty}^{-\Delta}\frac{dx}{2\pi}\frac{1}{x+y-\Gamma}\,\frac{1}{x-\Gamma}\right.\\
 &  &\ \ \ \ \ \ \ \ \ \ \ \ \ \ \ \ \ \ \ \ \ \ \ \ \ \ \ \ \ \ \ \ \ \ \ \ \ \ \ \ \  \left.+\int_{\Delta}^{-y}\frac{dx}{2\pi}\frac{1}{x+y-\Gamma}\,\frac{1}{x+\Gamma}+\int_{-y}^{\infty}\frac{dx}{2\pi}\frac{1}{x+y+\Gamma}\,\frac{1}{x+\Gamma}\right)\\
 &  & +U^{2}\left(W\left(\omega_{b}\right)+\int_{\Delta}^{\omega}\frac{dy}{2\pi}\frac{1}{\omega-y+\Gamma}\right) \left(W\left(-\omega_{b}\right)^{2}+W\left(\omega_{b}\right)^{2} +\int_{-\infty}^{-y}\frac{dx}{2\pi}\frac{1}{x+y-\Gamma}\,\frac{1}{x-\Gamma}\right.\\
 &  & \left.\ \ \ \ \ \ \ \ \ \ \ \ \ \ \ \ \ \ \ \ \ \ \ \ \ \ \ \ \ \ \ \ \ \ \ \ \ \ \ \ \ +\int_{-y}^{-\Delta}\frac{dx}{2\pi}\frac{1}{x+y+\Gamma}\,\frac{1}{x-\Gamma}+\int_{\Delta}^{\infty}\frac{dx}{2\pi}\frac{1}{x+y+\Gamma}\,\frac{1}{x+\Gamma}\right)\\
 &  & +U^{2}\left(\int_{\omega}^{\infty}\frac{dy}{2\pi}\frac{1}{\omega-y-\Gamma}\right) \left(W\left(-\omega_{b}\right)^{2}+W\left(\omega_{b}\right)^{2}+\int_{-\infty}^{-y}\frac{dx}{2\pi}\frac{1}{x+y-\Gamma}\,\frac{1}{x-\Gamma}\right.\\
 &  & \left.\ \ \ \ \ \ \ \ \ \ \ \ \ \ \ \ \ \ \ \ \ \ \ \ \ \ \ \ \ \ \ \ \ \ \ \ \ \ \ \  \ +\int_{-y}^{-\Delta}\frac{dx}{2\pi}\frac{1}{x+y+\Gamma}\,\frac{1}{x-\Gamma}  +\int_{\Delta}^{\infty}\frac{dx}{2\pi}\frac{1}{x+y+\Gamma}\,\frac{1}{x+\Gamma}\right)\\
\end{eqnarray*}
\end{widetext}
 As the states are all above and below the superconducting energy gap,
we consider the contribution of in-gap states separate from the integrals
and sum the weight of these states with the integral.  The following phase space diagram \ref{fig. integdomain} shows the domain of integration that is restricted to the grey regions on top of the red lines (Andreev bound states).  The energy $\omega_{b}$
is the bare bound-state energy that can be reached by setting $U=0$
in Eq. (\ref{eq. w}). In the second line the second parenthesis is written
considering negative $y$ and the remaining lines are for positive
$y$.  By the use of the condition $\Delta \ll \Gamma$ the above integration is simplified into

\begin{eqnarray*}
& & U^{2}\left(\int_{-\infty}^{0}\frac{dy}{2\pi}\frac{1}{\omega-y+\Gamma}\right)\left(\frac{2\Gamma\ln\frac{\Gamma-y}{\Gamma}}{-y(-y+2\Gamma)}\right)\\ & & +U^{2}\left(\int_{0}^{\omega}\frac{dy}{2\pi}\frac{1}{\omega-y+\Gamma}+\int_{\omega}^{\infty}\frac{dy}{2\pi}\frac{1}{\omega-y-\Gamma}\right)\left(\frac{2\Gamma\ln\frac{\Gamma+y}{\Gamma}}{y(y+2\Gamma)}\right)
\end{eqnarray*}

To simplify the analysis we considered the small energy-gap limit ($\frac{\Delta}{\Gamma}\ll1$)
and approximate $\Delta$ in the domain of integration to zero. Let
us first do the continuous part. After a change of variables from
$y$ to $-y$ in the first integral and with the change of variable
in all integrals into $s:=\frac{|y|(|y|+2\Gamma)}{\Gamma^{2}}$ for
positive values of $y$, the result is simplified to:

\begin{eqnarray*}
\frac{U^{2}}{8\pi^{2}} &  & \left(\int_{0}^{\infty}ds\left(\frac{1}{\omega+\Gamma\sqrt{1+s}}+\frac{1}{\omega-\Gamma\sqrt{1+s}}\right)\right.\\
 &  & +\int_{0}^{\frac{|\omega|(|\omega|+2\Gamma)}{\Gamma^{2}}}ds\left(\frac{1}{\omega-\Gamma\sqrt{1+s}+2\Gamma}-\frac{1}{\omega-\Gamma\sqrt{1+s}}\right) \\ && \ \ \ \ \ \ \ \ \ \ \ \ \ \ \ \ \ \ \ \ \ \times \left. \left(\frac{\ln\left(1+s\right)}{s\sqrt{1+s}}\right)\right)\\
\end{eqnarray*}

For small frequency $\omega$ we expand the integrands by the use
of $\frac{1}{1\pm x}\sim1\mp x$. This gives rise to:

\begin{eqnarray*}
\frac{U^{2}}{8\pi^{2}}& & \left(\frac{-2\omega}{\Gamma^{2}}\int_{0}^{\infty}ds\frac{\ln\left(1+s\right)}{s\left(1+s\right)^{\frac{3}{2}}}\right. 
\\ & & + \int_{0}^{\frac{|\omega|(|\omega|+2\Gamma)}{\Gamma^{2}}}ds\left(\frac{1}{\omega-\Gamma\sqrt{1+s}+2\Gamma}-\frac{1}{\omega-\Gamma\sqrt{1+s}}\right)
\\ && \ \ \ \ \ \ \ \ \ \ \ \ \ \ \ \ \ \ \ \ \ \times \left.\left( \frac{\ln\left(1+s\right)}{s\sqrt{1+s}}\right)\right)\\
\end{eqnarray*}

The first integral results into $\frac{1}{4}\left(\frac{U}{\pi\Gamma}\right)^{2}\left(4-\frac{\pi^{2}}{2}\right)\omega$.
The second integral gives $\omega$-linear term as for small frequency
$\omega$. To see this we first expand the integrand and simplify
it, then we take the integration that in the general case becomes
$2\ln(1+s)/\sqrt{1+s}+4/\sqrt{1+s}-\ln(1+\sqrt{1+s})\ln(1+s)-2 dilog(1+\sqrt{1+s})-2 dilog\sqrt{1+s}$
. If we expand this for small $s$  and calculate it in the integral
domain from 0 to $\frac{2\omega}{\Gamma}$ the result is proportional to $s$ and the corresponding self-energy becomes $\frac{1}{2}\left(\frac{U}{\pi\Gamma}\right)^{2}\omega$.
The total contribution of the continuum into the self-energy becomes
$\frac{1}{2}\left(\frac{U}{\pi\Gamma}\right)^{2}\left(3-\frac{\pi^{2}}{2}\right)\omega$.
The full self-energy contains both Andreev states on top of the continuum
limit. One can show that the weight of bound states for the case of
$\omega_{b}=\Delta\cos\frac{\varphi}{2},$ via Eq. (\ref{eq. w}), $W=\frac{\Delta}{\Gamma}|\sin\frac{\varphi}{2}|$
and $W_{\Delta}=\pm W.$ Consequently in the limit of $\frac{\Delta}{\Gamma}$
the discrete contributions cancel out and the self-energy is left
with its value at the continuum value:

\[ \Sigma\left(\omega\right)=-\frac{1}{2}\left(\frac{U}{\pi\Gamma}\right)^{2}\left(3-\frac{\pi^{2}}{4}\right)\omega\]

The two terms in Eq. (\ref{eq. GGGt}) can be written for opposite spins therefore the 1/2 factor is doubled.
\vspace{1cm}

\textbf{Diagram b)}

\begin{figure}[h]
 \includegraphics[width=7cm]{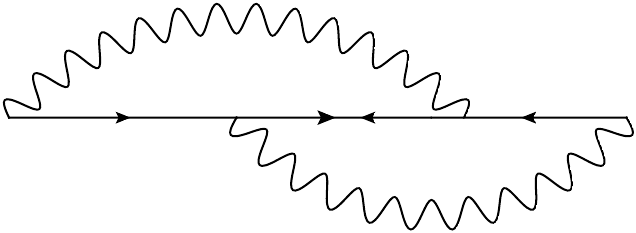}\\ 
\end{figure}

\[
\Sigma(\omega)  =  U^{2}\int \frac{dy}{2\pi} \int \frac{dx}{2\pi}\, G^{o}(y) F^{o}(\omega+x+y) G^{o}(x) \\
\]

In the leading this gives rise to 
\begin{widetext}

\begin{eqnarray*}
 U^{2} & & \left(  \int_{R-[-\Delta,\Delta]} \frac{dy}{2\pi} \int_{R-[-\Delta,\Delta]} \frac{dx}{2\pi}\  \frac{1}{y+sgn(y) \Gamma} \frac{\Delta \Gamma \cos \frac{\phi}{2}}{\left(\omega+x+y\right)\left(\omega+x+y+\Gamma\right)^{2}} \ \frac{1}{x+sgn(x)\Gamma}\right.\\ 
& & +  \sum_{j=+,-}W_{\Delta} (j |\omega_{b}|)  \int \frac{dy}{2\pi} \int \frac{dx}{2\pi} \frac{1}{x+sgn(x)\Gamma}\ \frac{1}{y+sgn(y)\Gamma} \delta\left(\omega+x+y-j|\omega_{b}|\right) \\
& &  \left. + 2 \sum_{j=+,-}W (j |\omega_{b}|)  \int \frac{dy}{2\pi} \int \frac{dx}{2\pi} \delta(x-j |\omega_{b|}) \frac{\Delta \Gamma \cos \frac{\phi}{2}}{\left(\omega+x+y\right)\left(\omega+x+y+\Gamma\right)^{2}}\ \frac{1}{y+sgn(y)\Gamma} + \cdots \right)  \\
\end{eqnarray*}
 \end{widetext}

 The third line is of second degree in $\Delta$ and we need only to compute the first two lines to represent the leading order. One can compute after applying the approximation $\ln (1+x) \approx x $ the second  term in the limit of $\Delta \ll \Gamma$, which gives rise to $ \frac{U^{2}}{4\pi^{2}} \sum_{j=+,-}W_{\Delta}(j |\omega_{b}|) \frac{4}{\Gamma}$ and because $W_{\Delta}(\omega)= -W_{\Delta}(-\omega)$, the outcome is zero. In other words, there is no $\sin \frac{\phi}{2}$ contribution into the off-diagonal self-energy of Anderson model coupled to superconductor leads.  The central integrand in the first line is expanded with respect to $\omega$. The term independent of $\omega$ is the only nonzero term for the limit of small $\omega$ because the term proportional to $\omega$ becomes an even function over the entire real numbers that is identical to zero. The only nonzero term  is
 
 \begin{widetext}

\begin{eqnarray*}
 \frac{U^{2}}{4 \pi^{2}}   \Delta \Gamma \cos \frac{\phi}{2} & & \left( \left( \int_{-\infty}^{-\Delta} dy \frac{1}{y- \Gamma} + \int_{\Delta}^{\infty} dy \frac{1}{y+\Gamma} \right) \right. \\ 
& & \left. \left(\int_{-\infty}^{-\Delta} dx  \frac{1}{\left(x+y\right)\left(x+y+\Gamma\right)^{2}} \ \frac{1}{x-\Gamma}+ \int_{\Delta}^{\infty}   dx \frac{1}{\left(x+y\right)\left(x+y+\Gamma\right)^{2}} \ \frac{1}{x+\Gamma}\right) \right) \\
\end{eqnarray*}
\end{widetext}

The result after taking the integration becomes:

\[
\Sigma_{\textup{off}}\approx \left(\frac{U}{\pi \Gamma}\right)^{2} \Delta  \cos \frac{\phi}{2} \left(  \frac{ \pi^{2}}{6} - 1\right)  
\]

%%%%%%%%%%%%%%%%%%%%%%%%%%%%%%%%%%%%%%%%%%%%%%%%%%%%%%%%%%%%%
%%%%%%%%%%%%%%% B %%%%%%%%%%%%%%%%%%%%%%%%%%%%%%%%%%%%%%%%%%%
%%%%%%%%%%%%%%%%%%%%%%%%%%%%%%%%%%%%%%%%%%%%%%%%%%%%%%%%%%%%%

\section{THE PARTICLE-HOLE ASYMMETRIC CASE}

To complete the discussion about the self-energy let us comment on
the asymmetric case. In general one can break the particle-hole symmetry.
In this case one should apply the modified self-energy computed from
the interpolation between the small and large $U$ in \cite{roderoSigma}
into the Dyson-Gorkov equation. This ends up with the following self-energy
modification: 
\begin{widetext}

\begin{equation}
\tilde{\Sigma}:=\frac{1}{1-\theta^{2}(\alpha^{2}\omega^{2}-\beta^{2}\Delta^{2})}\left(\begin{array}{cc}
\alpha\omega+\theta(\alpha^{2}\omega^{2}-\beta^{2}\Delta^{2}) & \beta\Delta\\
\beta\Delta & \alpha\omega-\theta(\alpha^{2}\omega^{2}-\beta^{2}\Delta^{2})\end{array}\right),
\label{eq. modSE}
\end{equation}
\end{widetext}

 where $\theta:=\eta/[U^{2}\langle n\rangle(1-\langle n\rangle)]$
which vanishes in the half-filling particle-hole symmetric case.

In this case, the bound states are still the solution to Eq. (\ref{eq. Eb})
if we replace $\alpha$ with $\tilde{\alpha}=\frac{\alpha}{1-\theta^{2}(\alpha^{2}\omega^{2}-\beta^{2}\Delta^{2})}$,
as well as $\beta$ with $\tilde{\beta}=\frac{\beta}{1-\theta^{2}(\alpha^{2}\omega^{2}-\beta^{2}\Delta^{2})}$
and also $\eta$ with $\tilde{\eta}=\frac{\theta(\alpha^{2}\omega^{2}-\beta^{2}\Delta^{2})}{1-\theta^{2}(\alpha^{2}\omega^{2}-\beta^{2}\Delta^{2})}$.

In the asymmetric case there are four solutions to the bound state
energy Eq. (\ref{eq. Eb}). These are produced from a splitting in
the two solutions obtained form the particle-hole symmetry. One can
try a bound-state $E_{b}\sim\sqrt{1-\tau\sin^{2}\phi}$ into the equation
and come up with an effective transmission coefficient that depends
on the Coulomb repulsion. 

\begin{equation}
\tau^{-1}=1+\theta\frac{\alpha^{2}\omega^{2}-\beta^{2}\Delta^{2}}{\Gamma-\Gamma\theta^{2}(\alpha^{2}\omega^{2}-\beta^{2}\Delta^{2})}.
\end{equation}

%%%%%%%%%%%%%%%%%%%%%%%%%%%%%%%%%%%%%%%%%%%%%%%%%%%%%%%%%%%%%
%%%%%%%%%%%%%%% B I B L I O G R A P H Y %%%%%%%%%%%%%%%%%%%%%
%%%%%%%%%%%%%%%%%%%%%%%%%%%%%%%%%%%%%%%%%%%%%%%%%%%%%%%%%%%%%

%let's give it a shot with Bibtex
%\bibliographystyle{plain}

%\bibliographystyle{prsty}
\bibliography{frankslibrary,frankspapers,MAlibrary}

\end{document}